\documentclass[11pt,tightenlines,eqsecnum,floats,aps,amsmath,amssymb,nofootinbib,prd,shownopacs,floatfix]{revtex4-2}

\usepackage{graphicx}
\usepackage{epstopdf}
\usepackage{latexsym}
\usepackage{amssymb}
\usepackage{amsmath}
\usepackage{color}
\usepackage{mathrsfs}
\usepackage{xparse}
\usepackage{float}
\usepackage{mathtools}

\usepackage[center]{subfigure}

\begin{document}
  \renewcommand\arraystretch{2}
 \newcommand{\bq}{\begin{equation}}
 \newcommand{\eq}{\end{equation}}
 \newcommand{\bqn}{\begin{eqnarray}}
 \newcommand{\eqn}{\end{eqnarray}}
 \newcommand{\nb}{\nonumber}
 \newcommand{\lb}{\label}
 \newcommand{\cb}{\color{blue}}
    \newcommand{\cc}{\color{cyan}}
        \newcommand{\cm}{\color{magenta}}
\newcommand{\rc}{\rho^{\scriptscriptstyle{\mathrm{I}}}_c}
\newcommand{\rd}{\rho^{\scriptscriptstyle{\mathrm{II}}}_c}
\NewDocumentCommand{\evalat}{sO{\big}mm}{%
  \IfBooleanTF{#1}
   {\mleft. #3 \mright|_{#4}}
   {#3#2|_{#4}}%
}
\newcommand{\PRL}{Phys. Rev. Lett.}
\newcommand{\PL}{Phys. Lett.}
\newcommand{\PR}{Phys. Rev.}
\newcommand{\CQG}{Class. Quantum Grav.}
\newcommand{\parallelsum}{\mathbin{\!/\mkern-5mu/\!}}

\title{Quantum gravity might restrict a cyclic evolution}

\author{Bao-Fei Li$^{1}$}
\email{baofeili1@lsu.edu}
\author{Parampreet Singh$^1$}
\email{psingh@lsu.edu}
\affiliation{
$^{1}$ Department of Physics and Astronomy, Louisiana State University, Baton Rouge, LA 70803, USA}


\begin{abstract}

It is generally expected that in a non-singular cosmological model a cyclic evolution is straightforward to obtain on introduction of a suitable choice of a scalar field with a negative potential or a negative cosmological constant which causes a recollapse at some time in the evolution. We present a counter example to this conventional wisdom. Working in the realm of loop cosmological models with non-perturbative quantum gravity modifications we show that a  modified version of standard loop quantum cosmology based on Thiemann's regularization of the Hamiltonian constraint while generically non-singular does not allow a cyclic evolution unless some highly restrictive conditions hold. Irrespective of the energy density of other matter fields, a recollapse and hence a cyclic evolution is only possible if one chooses an almost Planck sized negative potential of the scalar field or a negative cosmological constant. Further, cycles when present do not occur in the classical regime. Surprisingly, a necessary condition for a cyclic evolution, not singularity resolution, turns out to be a violation of the weak energy condition. These results are in a striking contrast to standard loop quantum cosmology where obtaining a recollapse at large volumes and a cyclic evolution is straightforward, and, there is no violation of weak energy condition. On one hand our work shows that some quantum cosmological models even though non-singular and bouncing are incompatible with a cyclic evolution, and on the other hand demonstrates that differences in various quantization prescriptions in loop cosmology need not be faint and buried in the pre-bounce regime,  but can be striking and profound even in the post-bounce regime.

\end{abstract}

\maketitle

\section{Introduction}
\label{introduction}
\renewcommand{\theequation}{1.\arabic{equation}}\setcounter{equation}{0}

Whether or not our universe goes through a cyclic evolution is a question which has intrigued both philosophers and scientists for a long time. The first rigorous construction of a cyclic cosmological model was presented by Tolman \cite{tolman} using Einstein's theory of general relativity (GR) for a universe with a positive spatial curvature. Since then, cyclic models in different avatars have been proposed (see for eg. \cite{steadystate,Lemaitre1933,bonnor,giao1963,Pachner1965,Narlikar1966,rosen1989,lqc1, lqc2,quasi-steadystate,steinhardt2002, apsv, varyingconstants, bp, greene, biswas, baum, kanekar-sahni-shtanov, sahni-toporensky, hysteresis1, hysteresis2, hysteresis3, lqc-cyclic1,lqc-cyclic2,lqc-cyclic3, lqc-cyclic4,lqc-inflation1,lqc-inflation2, rb-ekpyo,penrose,rb,brown}).  The first requirement to construct a viable model of cyclic evolution is a robust mechanism to resolve big bang/big crunch singularities. The second ingredient is some form of matter-energy or curvature to cause a recollapse. In construction of cyclic models, it is the first requirement, a non-singular bounce of the scale factor which turns out to be most challenging to meet, unless one sacrifices weak energy condition which results in a host of difficulties. To obtain a non-singular bounce often needs inputs from non-perturbative regime of quantum gravity and this is where models based on LQG stand out where resolution of strong curvature singularities turns out to be a generic phenomena for various isotropic and anisotropic models \cite{ps09,Singh:2011gp,Saini_2016,Saini_2017,Saini_2018}. The second requirement for cyclic models is generally considered easy to satisfy either with a positive spatial curvature as in Tolman's model or with a negative potential which can be constant or varying in some of the above models. Since recollapse generally occurs in a macroscopic classical universe, this requirement  is straightforward to meet even with a small magnitude of the negative potential.

The conventional wisdom in construction of cyclic models is that once the singularity is taken care of,  an appropriate choice of some matter-energy content or spatial curvature can be readily used to obtain a cyclic evolution. The goal of this manuscript is to present a counter-example to this understanding for a non-singular cosmological model. The premise of our work is loop quantum cosmology (LQC) \cite{as-status} which is a quantization of cosmological spacetimes following techniques of LQG.  Before going into the details of a particular version of LQC which does not favor a cyclic evolution, let us note some highlights of the non-singular dynamics in standard LQC.
Unlike the Wheeler-DeWitt quantum cosmology based on a continuum spacetime manifold, LQC is based on a discrete quantum geometry arising from the non-perturbative quantum geometric effects. These are responsible for a resolution of the big bang singularity replacing it with a big bounce at Planckian curvature scale \cite{aps1,aps2,aps3,slqc}. Backward evolution of a macroscopic universe towards the classical big bang singularity is well captured by GR up to a percent of the Planck curvature beyond which non-local quantum gravitational effects result in a significant departure from GR. Due to underlying quantum geometry, evolution in LQC is governed by a second order finite difference equation, but one can also extract an effective spacetime description using semi-classical states \cite{numlsu-2,numlsu-3} from which a modified Friedman dynamics with  quadratic modifications to energy density (with a negative sign) can be obtained \cite{ps06,aps3}. This additional term causes the Hubble rate to vanish at a maximum energy density determined by the quantum geometry and the universe reaches a pre-bounce regime where it soon becomes classical once again. In contrast to various other bouncing models, singularity resolution is obtained without any violation of weak energy conditions. In the absence of a potential, the resulting picture across the bounce in the standard LQC turns out to be symmetric. Further, consistent quantum probabilities for a big bounce turn out to be unity for arbitrary superposition of quantum wavefunctions \cite{craig-singh}. Notably, LQC dynamics turns out to be favorable for both inflationary and cyclic models. As examples, quantum bounce in LQC results in inflationary attractors for both isotropic and anisotropic spacetimes \cite{lqc-cyclic2, sloan1, sloan2, gupt-ps}, makes onset of inflation, in particular for low energy models,  in closed universes much easier as compared to the classical theory \cite{hysteresis1,hysteresis2, warm-inflation}, and alleviates problems with a non-singular turnaround of moduli field in the Ekpyrotic/cyclic models \cite{lqc-cyclic2, lqc-cyclic3}. As far as the question of a cyclic evolution is concerned, standard LQC allows a non-singular cyclic dynamics in the presence of spatial curvature \cite{apsv}, a negative cosmological constant \cite{bp}, Ekpyrotic potentials \cite{lqc-cyclic2,lqc-cyclic3, lqc-cyclic4} and certain types of clocks \cite{lqc-inflation1}.

Similar to any quantum cosmological model, LQC must deal with various quantization and regularization ambiguities to establish robustness of such predictions.  In LQC, one of such ambiguities arises in obtaining the quantum Hamiltonian constraint for spatially-flat isotropic models. In the quantization procedure one encounters two terms in the Hamiltonian constraint corresponding to the Euclidean and Lorentzian parts. In the spatially-flat model, one option is to utilize the symmetries of the cosmological spacetime to combine these terms before quantization \cite{abl}. This procedure results in the standard LQC. On the other hand, if one does not exploit this symmetry of the spatially-flat universe and treats Euclidean and Lorentzian terms independently during quantization procedure, following Thiemann's regularization of the Hamiltonian constraint in LQG,  one obtains a modified version of standard LQC, also known as mLQC-I in literature \cite{lsw2018}. This model was first analyzed in \cite{YDM09} and recently obtained following LQG techniques  by  computing the expectation values of the Hamiltonian constraint with the complexifier coherent states \cite{DL17}. Since this regularization aims to capture some of the more generic features of the quantization of cosmological spacetimes motivated from full LQG, it is pertinent to ask whether it yields similar predictions as standard LQC. This question has been partly answered in the context of singularity resolution and the inflationary paradigm. It turns out that LQC and mLQC-I both yield a generic resolution of strong curvature singularities \cite{saini-singh-mlqc-1} and are compatible with inflation \cite{lsw-inflation} resulting in an almost identical power spectrum at ultra-violet scales \cite{agullo, lsw2019a}. But, there are also some important differences. First, the evolution of the universe in mLQC-I is asymmetric with respect to the bounce. In the contracting phase, the universe  quickly tends to a quasi-de Sitter phase with an emergent cosmological constant of Planckian magnitude \cite{pawlowski,lsw2018} and a rescaled Newton's constant \cite{lsw2018}. The second difference arises in the modifications to the dynamical equations. At the quantum level, the difference equation is fourth order, and, at the effective spacetime level, the modified Friedmann equation has higher orders in energy density \cite{saini-singh-mlqc2,lsw2018}. Finally, there are significant differences in the behavior of the primordial power spectrum in the infra-red regime  which can potentially lead to signatures in non-gaussianities \cite{lsw2019a,lsw2019b}. Despite these differences, standard LQC and mLQC-I obtained from Thiemann's regularization of the Hamiltonian constraint have a good qualitative agreement with each other in the post-bounce regime for inflationary potentials and the compatibility with inflationary dynamics is a robust prediction of different versions of LQC.

Given that in the LQC version with Thiemann's regularization of the Hamiltonian constraint the evolution across the  bounce is highly asymmetric with one of the branches having a Planck curvature even at large volumes, it is pertinent to ask whether this modification of standard LQC results in a viable cyclic evolution.  The answer to this question is also related to a broader issue, that is  regardless of the quantization choices, whether the existence of the cyclic universes is a robust feature of the loop cosmologies. In order to address this problem, one has to start with  the qualitative behavior of the mLQC-I universe  when it is filled with some particular form of the matter content that can readily cause a cyclic universe in LQC. We consider the case when the matter content  only consists of one single scalar field and introduce a negative  potential, constant as well as varying,  to obtain a recollapse. Unlike in the presence of the positive potentials, some major qualitative differences between standard LQC and mLQC-I become apparent for negative potentials. These can also be viewed as differences between any conventional bouncing model which mimics features of standard LQC and the Thiemann regularized LQC. These differences are the following: (i) Unlike the  standard LQC where as in GR, the minimum possible energy density is zero, it is negative in mLQC-I given by $\rho_{\mathrm{min}} \approx -0.023$ in the Planck units. While this magnitude is approximately $2\%$ of the Planck density, it is $24\%$ of the maximum allowed energy density in mLQC-I. (ii) In the standard LQC, the recollapses in each cycle are qualitatively of the same type when the negative potential causes the minimum energy density to be achieved. There is a single branch extending from pre-bounce to post-bounce eras described by the same modified Friedmann equation. But in mLQC-I, because of the richer regularization of the Hamiltonian constraint there are two branches in evolution joined at the bounce. These are  labelled as $b_-$ and $b_+$ branches following \cite{lsw2018} and correspond to different modifications to the Friedmann equation. The $b_-$ branch leads to classical GR, but the $b_+$ branch is purely quantum gravitational in character with
an emergent Planck-sized cosmological constant.
These branches lead to two types of recollapse points alternating each other. One recollapse occurs at the vanishing energy density (in the $b_-$ branch). This recollapse is thus classical in nature. The other recollapse occurs at $\rho_{\mathrm{min}}$ (in the $b_+$ branch) and is thus quantum gravitational in nature. For a cyclic evolution, the universe must pass through both the classical and quantum recollapse points alternately. (iii) For a recollapse in $b_+$ branch to occur, the minimal of the potential must be more negative than $\rho_{\mathrm{min}}$. Because of this constraint, a very large parameter space of negative potentials including negative cosmological constants which can easily cause a recollapse in the classical regime  in LQC and thus a cyclic evolution gets ruled out. The recollapse requires a negative potential which is almost Planckian in magnitude. (iv) In standard LQC, there is no violation of weak energy condition in presence of a negative potential. In mLQC-I, weak energy condition must be violated at the recollapse point in the $b_+$ branch.
Due to these differences we find that  unlike standard LQC, a cyclic evolution in mLQC-I is highly restrictive and only occurs in very special situations requiring an almost Planckian sized negative potential. In the cases where this cyclic evolution is possible, the energy density of the universe retains its Planckian character and does not reach a sustained period of a classical cyclic evolution.

This manuscript is organized as follows. In Sec. \ref{sec:analytical analysis}, using the effective dynamics of LQC and mLQC-I, we study the effects of a negative potential by focusing on the evolution of the energy density and $\sin(\lambda b)$ in each model. The emphasis will be placed on the necessary conditions for a negative potential to generate a cyclic universe in each model and the distinctions between the resulting cyclic universes.   In Sec. \ref{numerical}, based on the numerical results of the Hamilton's equations, we analyze the dynamics of the universe which is filled with a single scalar field with a negative cosmological constant and an Ekpyrotic potential. We study two types of cases. In the first situation we consider ordinary values of negative cosmological constant and the magnitude of the Ekpyrotic potential and show that cyclic evolution is not possible. We then study the cases where cyclic evolution is allowed albeit with an almost Planckian magnitude of the cosmological constant and the Ekpyrotic potential.  We summarize our main findings in Sec. \ref{summary}.
In the following, we use Planck units with $\hbar=c=1$ while keeping Newton's constant $G$ explicit in our formulae. In the numerical results, $G$ is also set to unity.

\section{Possibility of cyclic evolution in spatially-flat LQC and mLQC-I}
\label{sec:analytical analysis}
\renewcommand{\theequation}{2.\arabic{equation}}\setcounter{equation}{0}

In this section we first discuss the successive evolutionary stages of a spatially flat FLRW universe in LQC when a scalar field minimally coupled to gravity is considered. Using effective dynamics, whose validity for the sharply peaked semiclassical states  has been extensively corroborated via numerical simulations  for both isotropic and anisotropic spacetimes \cite{numlsu-1,numlsu-2,numlsu-3,numlsu-4}, we note some distinctive properties in presence of a negative potential in contrast to the positive potential. The latter has been  applied in detail to  study the phenomenological implications of quantum gravity in the early universe, especially on the inflationary spacetimes and quantum perturbations around the background quantum geometry \cite{agullo-singh}. In the following discussion the negative potential may also be a constant potential which mimics a negative cosmological constant or an Ekpyrotic/cyclic potential which for an appropriate choice of parameters results in a recollapse of a classical universe at macroscopic scales. Here  we focus on the change in $\sin(\lambda b)$ and the energy density as the universe evolves forward and backward in time. Then, we study the same using the effective dynamics in mLQC-I and highlight some important features of dynamics with a negative potential and contrast them with the situation in LQC.

\subsection{Characteristic evolution in LQC}
 The fundamental variables for loop quantization are the holonomy of the Ashtekar-Barbero connection and  the flux of the densitized triad, which in a spatially flat FLRW universe are symmetry reduced to a canonical pair $c$ and $p$, with $p$ related with the scale factor of the universe via $|p|=a^2$ and $c$ proportional to the time derivative of the scale factor in the classical theory. In the $\bar \mu$ scheme \cite{aps3} in which a quantum bounce generically takes place at a fixed Planck-scale energy density, it is more convenient to employ an equivalent set of canonical variables which are defined via $v=|p|^{3/2}$ and $b=c|p|^{-1/2}$ with their fundamental Poisson bracket given by $\{b, v\}=4\pi G\gamma$, where the Barbero-Immirzi parameter  $\gamma$ is usually chosen to be $0.2375$ based on the black hole thermodynamics in LQG. Moreover, if we consider a spatially flat FLRW universe filled with a single scalar field,  the matter sector consists of a scalar field $\phi$ and its conjugate momentum $p_\phi$ with the Poisson bracket $\{\phi, p_\phi\}=1$.

In terms of the canonical variables introduced above, the effective Hamiltonian constraint in LQC takes the form \cite{aps3}
\bq
\lb{Hamiltonian in LQC}
\mathcal {H}_\mathrm{LQC}=-\frac{3v \sin^2(\lambda b)}{8\pi G \gamma^2 \lambda^2}+\mathcal H_M,
\eq
here $\mathcal H_M$ stands for the matter sector of the constraint, which consists of the kinetic and potential energy of the scalar field and thus is explicitly given by
\bq
\lb{matter Hamiltonian}
\mathcal H_M=\frac{p^2_\phi}{2v}+ v\, U,
\eq
where $U$ denotes the potential energy of the scalar field. The Hamilton's equations for the effective dynamics of LQC read
\bqn
\lb{LQCvb}
\dot v&=&\frac{3 v}{2 \gamma \lambda} \sin(2\lambda b),\quad \quad
\dot b=-4\pi G \gamma \left( \rho+ P\right),\\
\lb{matter equations of motion}
\dot \phi&=&\frac{p_\phi}{v}, \quad \quad \quad  \dot p_\phi=-v \, U_{,\phi},
\eqn
where $U_{,\phi}$ denotes the derivative of the potential with respect to the scalar field. In addition,  the energy density and the pressure of the scalar field are given explicitly by
\bq
\label{energy density and pressure}
\rho=\frac{p^2_\phi}{2v^2}+U, \quad \quad P=\frac{p^2_\phi}{2v^2}-U,
\eq
so that the equation of state of the scalar field is defined by
\bq
\label{eos}
w=\frac{P}{\rho}.
\eq
From the vanishing of the total Hamiltonian constraint, one can also relate the energy density with the momentum $b$ via
\bq
\lb{energy density in LQC}
\rho=\frac{3\sin^2\left(\lambda b\right)}{8\pi G\gamma^2\lambda^2},
\eq
which implies  the energy density in LQC is always non-negative and lies in the range $\rho\in[0, \rho^\mathrm{LQC}_\mathrm{max}]$, with $\rho^\mathrm{LQC}_\mathrm{max}=\frac{3}{8\pi G \gamma^2 \lambda^2}$. From Eqs. (\ref{LQCvb}) and (\ref{energy density in LQC}), it is straightforward to arrive at the Friedmann equation in LQC, which reads
 \bq
 \lb{FriedmannLQC}
 H^2=\frac{8\pi G}{3}\rho\left(1-\frac{\rho}{\rho^\mathrm{LQC}_\mathrm{max}}\right).
 \eq
 As a result, the Hubble rate vanishes at $\rho=0$ or $\rho=\rho^\mathrm{LQC}_\mathrm{max}$. By computing $\ddot a $, one finds that the bounce happens at $\rho=\rho^\mathrm{LQC}_\mathrm{max}$ while $\rho=0$ corresponds to a recollapse point. Moreover, since only a scalar field with the energy density and pressure as given in (\ref{energy density and pressure}) is considered, $b$ decreases monotonically in the forward evolution of the universe as can been seen from the $\dot b$ equation in (\ref{LQCvb}). As a result, $b$ can also be used as a geometric clock to unfold the background evolution of the universe.

\begin{figure}
\includegraphics[width=8cm]{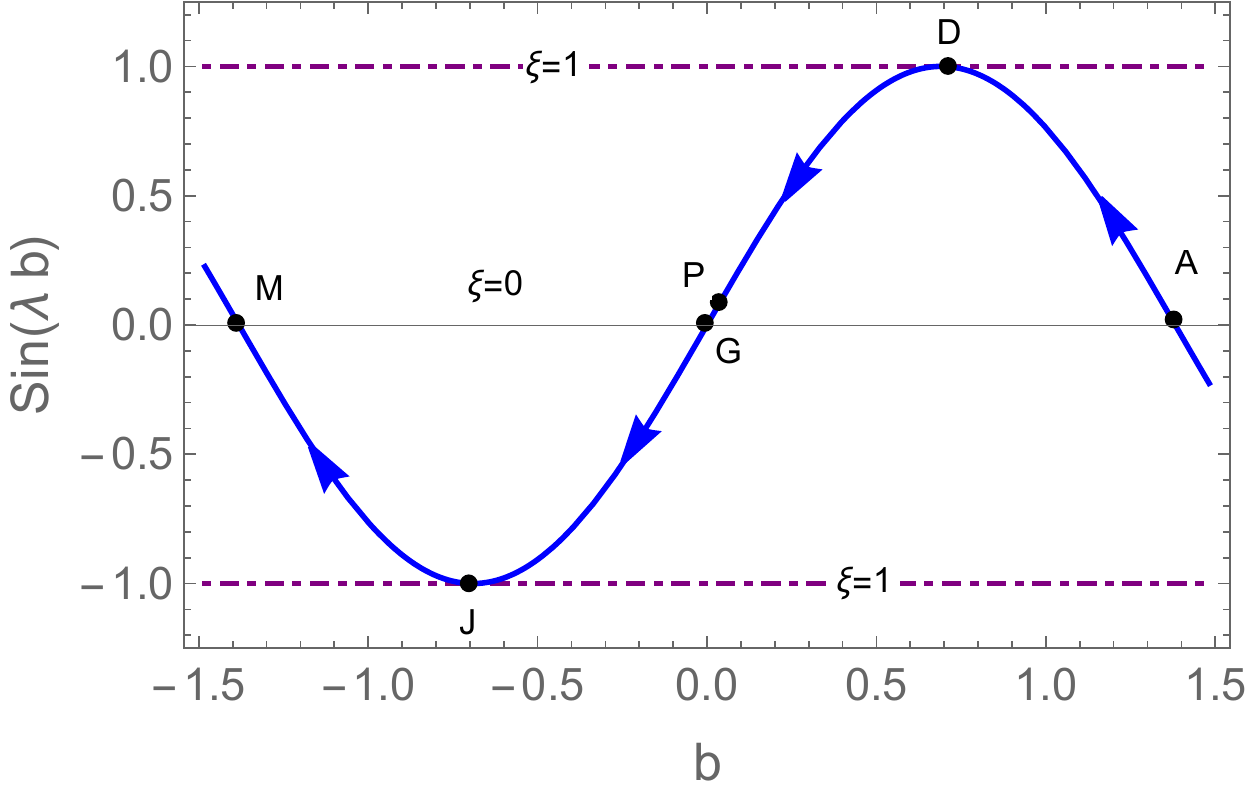}
\includegraphics[width=8cm]{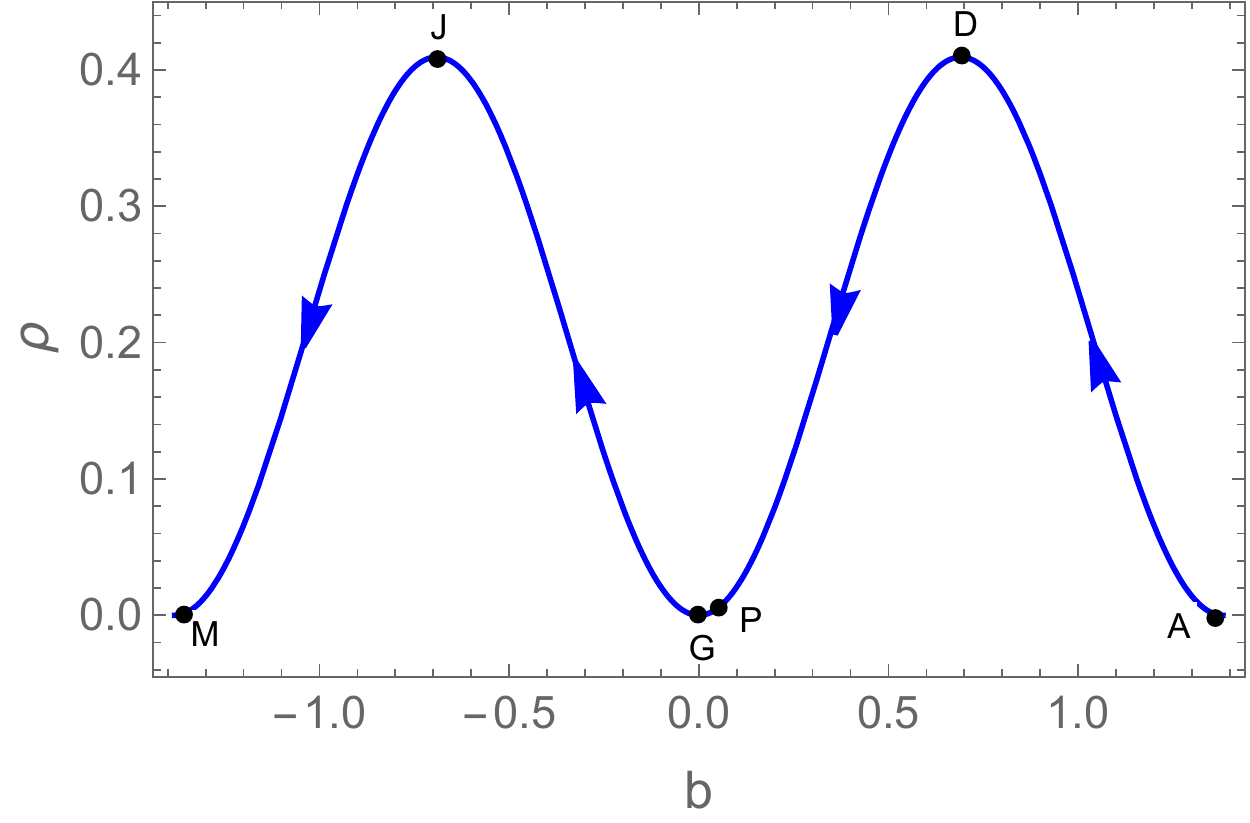}
\caption{In the figure,  $\sin(\lambda b)$ and $\rho$ are depicted in the forward evolution of the universe in LQC. The arrows denote the direction of time flow. The bounce points are labelled by $D$, $J$, and the recollapse points are labelled by $A$, $G$, $M$. $\xi$ stands for the ratio of the energy density of the scalar field over the maximum energy density $\rho^\mathrm{LQC}_\mathrm{max}$ in LQC. Point $P$ which is close to the recollapse point $G$ stands for the macroscopic universe representing the present epoch in this cyclic evolution.}
\label{LQC}
\end{figure}

 In Fig. \ref{LQC}, we show explicitly how the energy density of the universe evolves with respect to the clock $b$. Starting with our present macroscopic universe denoted by the point $P$, the universe can go through different stages depending on the sign of the potential of the scalar field.  For a positive potential, in the backward evolution, the universe starting at point $P$  reaches the bounce point $D$ and then tends towards point $A$. Note that in the forward evolution, from $D$ to $P$, the universe is in an expanding phase while from $A$ to $D$, the universe is in a contracting phase. However, with a positive potential, the universe can never reach the recollapse point $G$ in the future and the recollapse point $A$ in the past since the energy density in (\ref{energy density and pressure}) would not vanish completely for any non-zero positive potential. As a result,  there is no cyclic evolution of the universe when the scalar potential is positive. The evolution only lies between the boundary points $A$ and $G$.

 In case of a negative potential there are some changes from the evolution in case of the positive potential. With a cyclic (negative) potential, in the backward evolution, the universe would go through the bounce at point $D$ and reach the recollapse point $A$ in the past. And, in the forward evolution starting from the point $P$, the universe can reach the recollapse point $G$ when the kinetic energy of the scalar field is exactly cancelled by its potential energy and then enters into a contracting phase between $G$ and $J$, where at point $J$, a second bounce happens.  Afterwards, the universe undergoes expansion from $J$ to $M$. Thus, the evolution from $A$ to $M$ represents one complete cycle of the evolution. In the right panel of Fig. \ref{LQC}, we show the change in the energy density in the same cycle of evolution from point $A$ to point $M$. The universe would continue to evolve after reaching points $M$ and $A$. In this way, a cyclic evolution of the universe can be realized with a negative potential in LQC. It should be noted that in LQC, the energy density is always non-negative, that is, the weak energy condition is always satisfied irrespective of the sign of the potential.

 \subsection{Characteristic evolution in mLQC-I}

In mLQC-I, the fundamental canonical variables are the same as in LQC with the same Poisson brackets. The difference from LQC lies in the treatment of Lorentzian term in the Hamiltonian constraint where it is quantized independently rather than expressed in terms of the Euclidean term. Assuming that the effective description is still valid for the sharply-peaked quantum states, the effective Hamiltonian constraint in mLQC-I (denoted with ${H}_{\mathrm{I}}$) reads  \cite{YDM09,lsw2018}
\bq
\lb{Hamiltonian in mLQC-I}
\mathcal {H}_{\mathrm{I}} =\frac{3v}{8\pi G\lambda^2}\left\{\sin^2(\lambda b)-\frac{(\gamma^2+1)\sin^2(2\lambda b)}{4\gamma^2}\right\} +\mathcal{H}_M ~.
\eq
For a scalar field with a general potential term, the matter Hamiltonian $\mathcal{H}_M$ is still given by (\ref{matter Hamiltonian}).
From (\ref{Hamiltonian in mLQC-I}),  it is straightforward to derive the Hamilton's equations in mLQC-I which read,
\bqn
\lb{eomA}
\dot v&=&\frac{3v\sin(2\lambda b)}{2\gamma \lambda}\Big\{(\gamma^2+1)\cos(2\lambda b)-\gamma^2\Big\}, \quad \quad\dot b=-4\pi G \gamma \left( \rho+ P\right),\\
\lb{eomB}
\dot \phi&=&\frac{p_\phi}{v}, \quad \quad \quad  \dot p_\phi=-v U_{,\phi}.
\eqn
Since the total Hamiltonian constraint vanishes, one can  relate the energy density with the momentum $b$ in mLQC-I via
\bq
\lb{2.2}
\rho=-\frac{3\sin^2(\lambda b)}{8\pi G \gamma^2 \lambda^2}\Big\{\gamma^2 \sin^2\left(\lambda b\right)-\cos^2\left(\lambda b\right)\Big\},
\eq
 from which one can easily find the maximal and minimal energy densities in mLQC-I. More specifically, when $\sin^2(\lambda b)=\frac{1}{2(1+\gamma^2)}$, the energy density reaches its maximum value  $\rho_\mathrm{max}$, and when  $\sin^2(\lambda b)=1$, the energy density reaches its minimum value $\rho_\mathrm{min}$. These values are given by
 \bq
 \rho_\mathrm{max} =\frac{3}{32\pi G\gamma^2\lambda^2 (1+\gamma^2)} ~~ \mathrm{and} ~~  \rho_\mathrm{min}=-\frac{3}{8\pi G\lambda^2} .
 \eq

\begin{figure}
\includegraphics[width=8cm]{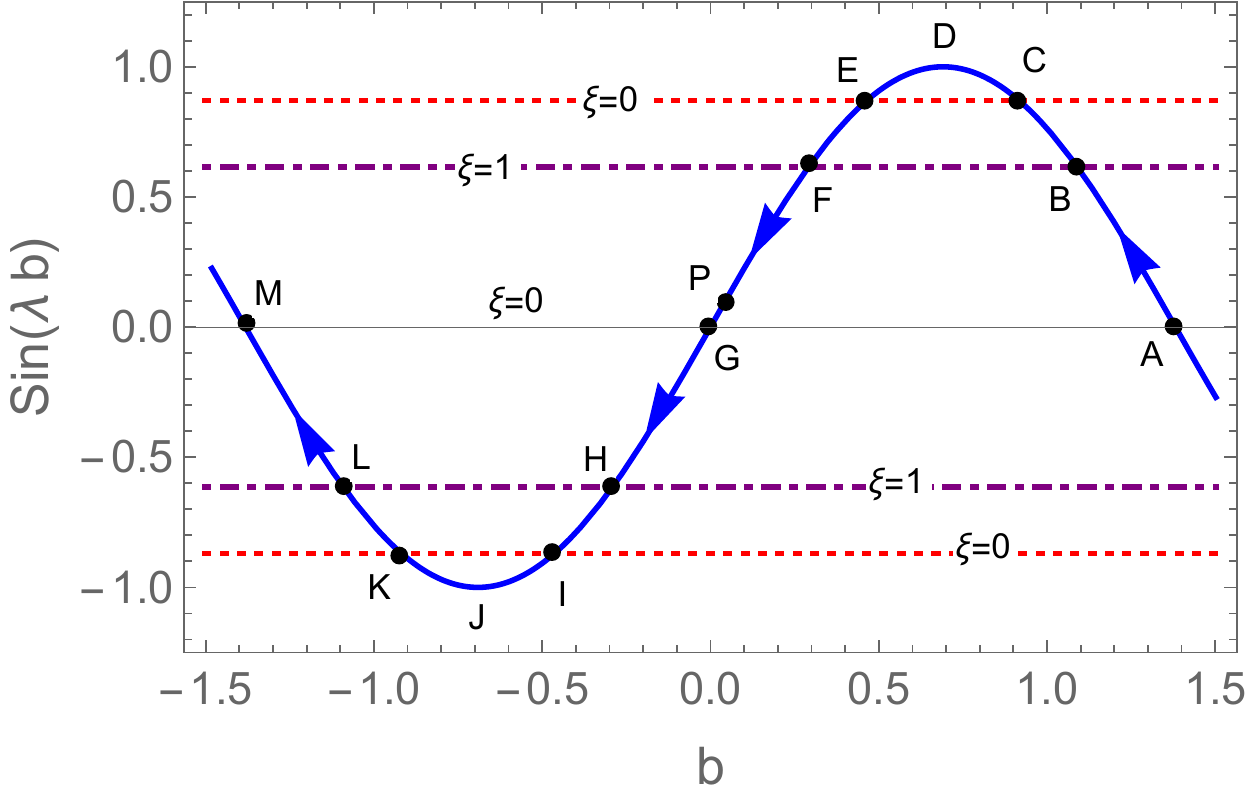}
\includegraphics[width=8cm]{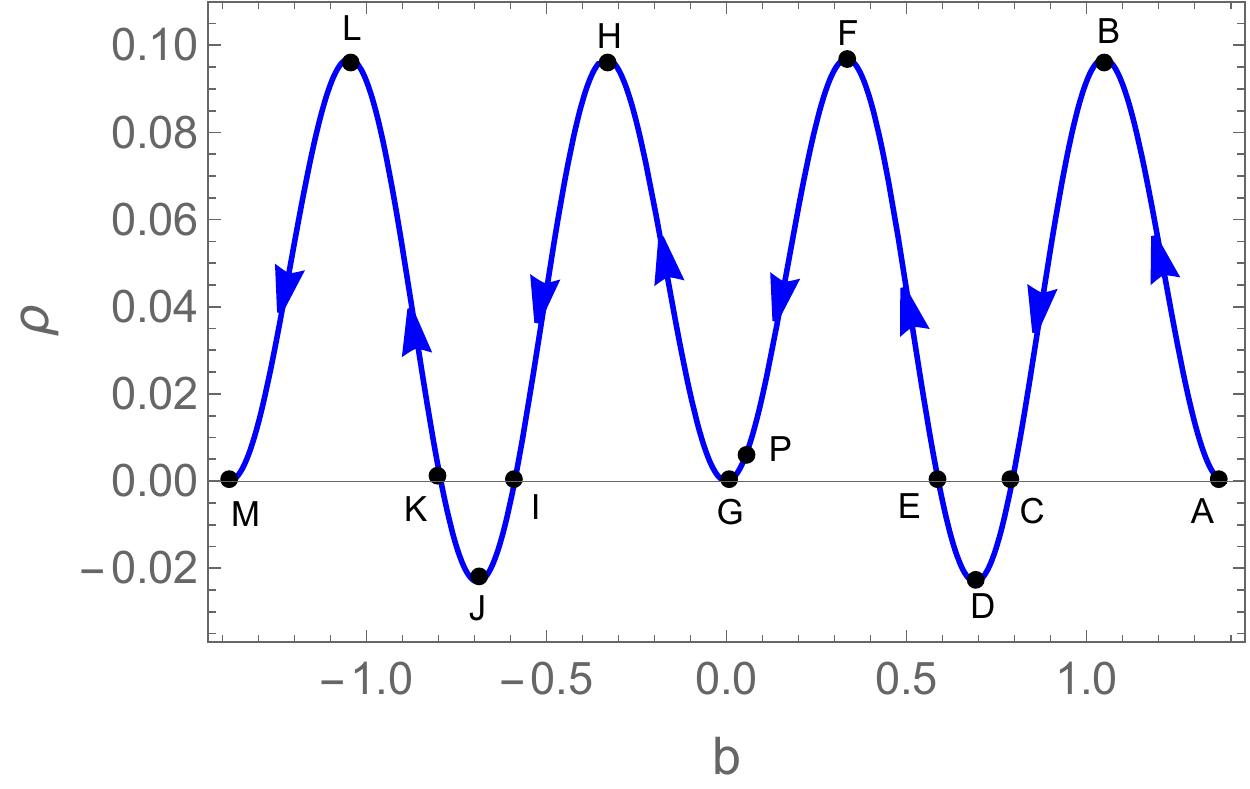}
\caption{In the figure, we  show the values of $\sin(\lambda b)$ and $\rho$ in mLQC-I when the universe evolves in the direction indicated by the arrows.  In the left panel,  $\xi$ stands for  the ratio of the energy density over the maximum energy density in mLQC-I. Hence, the line $\xi=0$ corresponds to the moment when the energy density vanishes and the line $\xi=1$ to the moment when the energy density reaches its maximum.  
The $b_+$ branches lie between points $B-F$ and $H-L$, and the rest of trajectory lies in $b_-$ branch.
In the right panel, we locate the same bounce and recollapse points in the $\rho$ vs $b$ plot which shows the matter energy density at each corresponding point.}
\label{mLQC-I}
\end{figure}

One of the characteristic properties of mLQC-I is that the evolution of the universe is asymmetrical  with respect to the bounce even when the potential vanishes identically. This becomes evident from the modified Friedmann equations which can be derived by using the $\dot v$ equation in (\ref{eomA}) and the relation (\ref{2.2}). It turns out that there exist two distinct branches which we  call $b_+$ branch and $b_-$ branch, following \cite{lsw2018}.  The modified Friedmann equation in the $b_-$ branch takes the form
\bq
\lb{b-}
H^2_-=\frac{8\pi G \rho}{3}\left(1-\frac{\rho}{\rho_\mathrm{max}}\right)\Bigg[1  +\frac{\gamma^2}{\gamma^2+1}\left(\frac{\sqrt{\rho/\rho_\mathrm{max}}}{1 +\sqrt{1-\rho/\rho_\mathrm{max}}}\right)^2\Bigg],
\eq
which, by requiring $H^2_-\ge0$, implies the energy density ranges over $\rho \in \big[0,\rho_\mathrm{max}\big]$ in the $b_-$ branch.  On the other hand, in the $b_+$ branch, the modified Friedmann equation reads
\bq
\lb{b+}
H^2_+=\frac{8\pi G\alpha  \rho_\Lambda}{3}\left(1-\frac{\rho}{\rho_\mathrm{max}}\right)\left[1+\left(\frac{1-2\gamma^2+\sqrt{1-\rho/\rho_\mathrm{max}}}{4\gamma^2\left(1+\sqrt{1-\rho/\rho_\mathrm{max}}\right)}\right)\frac{\rho}{\rho_\mathrm{max}}\right],
 \eq
 with $\alpha\equiv ({1-5\gamma^2})/({\gamma^2+1})$ and $\rho_\Lambda \equiv{3}/[{8\pi G\alpha \lambda^2(1+\gamma^2)^2}]$. For the $b_+$ branch, requiring $H^2_+\ge 0$ leads to $\rho \in[\rho_\mathrm{min}, \rho_\mathrm{max}]$ with $\rho_\mathrm{min}=-3/(8\pi G\lambda^2)\approx-0.023$. It should be noted that unlike in the $b_-$ branch where the Hubble rate vanishes at $\rho=0$ and $\rho=\rho_\mathrm{max}$, in the $b_+$ branch, the Hubble rate vanishes at $\rho=\rho_\mathrm{max}$ and $\rho=\rho_\mathrm{min}$ with $\rho_\mathrm{min}$ obtained from requiring the terms in the square bracket in \eqref{b+} vanish.
 For $\rho\rightarrow 0$, there emerges a rescaled Newton's constant $G_\alpha=\alpha G$ and a Planck-scale cosmological constant $ \rho_\Lambda=0.030$ \cite{lsw2018}, which is larger than the magnitude of the minimum energy density as $|\rho_\mathrm{min}|\approx0.023$.  The  $b_+$ branch and $b_-$ branch are connected via a quantum bounce that takes place at the maximum energy density $\rho_\mathrm{max}$.
 Similar to the case in LQC, for a single scalar field with the energy density and pressure given by (\ref{energy density and pressure}), the $\dot b$ equation in (\ref{eomA}) implies $b$ monotonically decreases as the universe evolves forward in time. As a result, $b$ also serves as a good global geometrical clock to unfold the dynamical evolution of the universe in mLQC-I. \\

 \noindent
 {\bf{Remark:}} As we show in the following, the universe in mLQC-I model can undergo a recollapse in two situations. One when energy density vanishes and the other when energy density becomes equal to $\rho_{\mathrm{min}}$. The first type of recollapse occurs in the $b_-$ branch and corresponds to points $A,  G,  M$ in Fig. 2. These recollapse points take place at the vanishing energy density and thus are the classical recollapse points.  The second type of recollapse occurs in the $b_+$ branch when the terms in the square bracket of \eqref{b+} vanish, which correspond to points $D$ and $J$ in Fig. 2.
 These recollapse points can only take place at the negative energy density $\rho=\rho_\mathrm{min}$ with an almost Planckian magnitude. Hence they are recollapse point originating from the quantum gravitational effects. As one can see from Fig. 2, for a universe to undergo cyclic evolution in mLQC-I it must undergo both kinds of recollapses and hence there is a violation of weak energy condition due to recollapse at a negative $\rho_{\mathrm{min}}$ in the $b_+$ branch. In contrast, in LQC, there is no such violation of weak energy condition at the recollapse point and hence for a cyclic evolution. Finally, note that in various cyclic models in cosmology generally a violation of weak energy condition is needed for resolution of the big bang singularity and a bounce. This is not the case for LQC as well as mLQC-I. \\

In Fig. \ref{mLQC-I}, we plot the evolution of  $\sin(\lambda b)$ and the energy density as the universe evolves in the time flow indicated by the arrows in the plots. In the left panel, the regime sandwiched between  two dot-dashed purple lines ($\xi=1$) corresponds to the $b_-$ branch, while the rest part of the solid blue curve belongs to the $b_+$ branch. Starting from the present macroscopic universe located at $P$, it undergoes various stages in the forward and backward evolution.
In the forward evolution, the recollapse point $G$ can only be reached when the potential of the scalar field becomes negative and $\rho$ becomes zero. This corresponds to a recollapse of the first kind as discussed in the {\it Remark} above. Afterwards, in the forward evolution from $G$ to $H$, the universe is in a contracting phase. At point $H$, the bounce takes place and the universe enters into the $b_+$ branch. The universe expands from $H$ to $J$ and contracts from $J$ to $L$. At points $I$ and $K$, the energy density becomes zero and an effective cosmological constant $\rho_\Lambda$ emerges leading to a rapid expansion of the universe, which implies  the universe is in a state of exponential expansion (near $I$) or contraction (near $K$). The universe undergoes a recollapse (of the second kind in above discussion) at point $J$. It should be noted that in the regime $I\rightarrow J\rightarrow K$, the energy density of the universe is negative which is in violation of the weak energy condition and even with a Planck-sized emergent cosmological constant the universe is not in a de Sitter phase. Rather the expansion in this phase is super-exponential or phantom universe like evolution (see for eg. \cite{ps-sami-dadhich}).  Finally, the universe re-enters into the $b_-$ branch when it is expanding from $L$ to $M$ and the cycle continues.

In the backward evolution,  the macroscopic universe at point $P$ would first contract to the bounce point $F$ and then expand to the point  $E$. Similar to the point $G$, the point $E$ can only be reached when the potential of the scalar field is negative. The backward evolution from $F$ to $B$ lies in the $b_+$ branch. At points $E$  and $C$, the energy density vanishes accompanied by an emergent cosmological constant $\rho_\Lambda$. During the backward evolution $E\rightarrow D\rightarrow C$, the energy  density of the universe turns out to be negative which violates the weak energy condition once again.  The evolution from $A$ to $M$ forms a complete cycle as shown in the left panel of Fig. \ref{mLQC-I} in presence of a suitable matter such as a negative cosmological constant or an Ekpyrotic potential. Note that for a cyclic evolution not only energy density must vanish but also become negative whose magnitude is approximately equal to 0.023 in Planck units. Thus, cyclic evolution in mLQC-I requires an almost Planckian valued negative potential.

To summarize the discussion in this subsection let us consider the right panel of Fig. \ref{mLQC-I} which helps us identify the locations of the bounces and recollapses when the universe evolves through various stages. All the local maxima in the $\rho$ vs $b$ plot, such as $B$, $F$, $H$ and $L$ are bounce points while all the local minima, such as $A$, $D$, $G$, $J$ and $M$ are recollapse points.  From the right panel, one can also easily identify the regimes in which the energy density becomes negative. Since the momentum $b$ has to continuously and monotonically evolve in a universe filled with a single scalar field, there are some necessary conditions for the generation of such universe in mLQC-I.  For example, if the potential of the scalar field is positive definite and the universe starts from the bounce  point $F$, then in the forward (backward) evolution, the universe can not reach point $G$ ($E$) in a finite period of time since $E$ and $G$ correspond to the universe with zero energy density which requires infinite volume when the potential of the scalar field is positive definite. In addition to introducing a negative potential, the necessary condition to have a cyclic universe in mLQC-I turns out to be more restrictive than in LQC. In LQC, one only needs a negative potential to ensure the occurrence of the recollapse, the total energy density is always non-negative as can be seen from the right panel of Fig. \ref{LQC}. However, in mLQC-I, the total energy density must be able to decrease to $\rho_\mathrm{min}$ which is a Planck-sized energy density. Thus, the weak energy condition must be violated in the cyclic evolution of the universe in mLQC-I.

\section{Numerical results on the possibility of cyclic evolution of a spatially flat FLRW universe in LQC and mLQC-I}
\label{numerical}
\renewcommand{\theequation}{3.\arabic{equation}}\setcounter{equation}{0}
In this section, we analyze some representative numerical solutions in  mLQC-I and LQC for the matter content consisting of a scalar field with a negative cosmological constant and  an Ekpyrotic potential and  discuss the scenarios in which a cyclic evolution is possible. As discussed in the previous section, in order to obtain a cyclic evolution in mLQC-I one needs an almost Planck sized value of a negative  potential and a violation of the weak energy condition. This is confirmed by the numerical investigations discussed below. The numerical analysis is based on the Hamilton's equations  in each model and thus the initial parameter space is composed of four phase space variables $v$, $b$, $\phi$ and $p_\phi$.  The numerical solutions are obtained using Mathematica and we use the Planck units and set $G$ equal to unity. In the numerical results, we show the evolution of several characteristic quantities for some representative initial conditions, such as volume and its conjugate variable and energy density. Another quantity of interest is the equation of state but since it becomes infinite at each recollapse points where energy density vanishes, we plot the inverse of the equation of state $w^{-1}$ to explicitly show the regions where the weak energy condition is violated with  $w^{-1}\in(0,-1)$ or equivalently $w\in(-1,-\infty)$.

 \subsection{The effective dynamics with a constant negative potential}

 When the potential $U$ is independent of the scalar field, it is equivalent to a cosmological constant, which can be set to
\bq
\lb{3a1}
U=\frac{\Lambda}{8\pi G},
\eq
where $\Lambda$ is the cosmological constant and the factor in the  denominator  is to recover the standard Friedmann equation with a cosmological constant in the classical limit.
In this case, the initial conditions are chosen right at the bounce where the energy density reaches its maximum. Since the Hamilton's equations in both LQC and mLQC-I are form invariant under the rescaling $v\rightarrow \beta v$ and $p_\phi \rightarrow \beta p_\phi$ with $\beta$ being any arbitrary constant, without loss of generality, the initial volume at the bounce can be set to  unity. Then, once the cosmological constant $\Lambda$ is specified, the momentum of the scalar field is given by
\bq
\lb{3.1}
p_{\phi_i}=\pm \sqrt{2\left(\rho^\mathrm{bounce}_\mathrm{max}-U_i\right)},
\eq
here $\rho^\mathrm{bounce}_\mathrm{max}$ refers to the maximum energy density at the bounce in LQC or mLQC-I and the subscript `i'  refers to quantities at the initial time $t_i=0$. In addition, in mLQC-I,  the momentum $b$ can be solved from (\ref{2.2}), yielding multiple values of the initial $b_i$ which correspond to different bounce points as shown in the left panel of Fig. \ref{mLQC-I}, such as $B$, $F$, $H$, $L$ etc. To be concrete, we  set the initial conditions  at point $F$. On the other hand, for LQC, the initial conditions are chosen at the bounce point $D$ in Fig. \ref{LQC}  where $b_i=\pi/(2\lambda)$ and the energy density reaches its maximum value.  Finally, in the case of a constant potential, the momentum of the scalar field becomes a  constant of motion  which implies that the scalar field can be used as a good global clock. As a result, the initial  value of the scalar field can be chosen arbitrarily. For  concreteness, in the numerical analysis, we choose $\phi_i=0$ at the initial bounce $t_i=0$. In the following, depending on the magnitude of the negative cosmological constant, two phenomenologically distinct  cases are studied. We first discuss the case when negative cosmological constant is such that its energy density is greater than $\rho_{\mathrm{min}} \approx -0.023$ (in Planck units) in mLQC-I. In such a case we show that no cyclic evolution is possible. We then consider a more negative cosmological constant with an almost Planckian magnitude such that $|\rho_\Lambda| > |\rho_{\mathrm{min}}|$ which allows a cyclic evolution.

\subsubsection{Non-cyclic evolution in mLQC-I}
When the energy density of the scalar field with a constant negative potential is larger than $\rho_\mathrm{min}=-3/(8\pi G\lambda^2)\approx -0.023$ there is no cyclic evolution.
 As an example of such a case, we choose the negative cosmological constant as
\bq
\lb{initial1}
\Lambda=-0.01,
\eq
leading to  $U_i=-3.98\times10^{-4}$ which is larger than the minimum energy density in mLQC-I, namely  $\rho_\mathrm{min}=-3/(8\pi G\lambda^2)\approx -0.023$. Using \eqref{3.1} the momentum of the scalar field equals  $0.44$ in mLQC-I and $0.91$ in LQC. Here, we take the positive sign of the momentum so that the value of the scalar field increases in the forward evolution of the cosmic time.

\begin{figure} [tbh!]
\includegraphics[width=7cm]{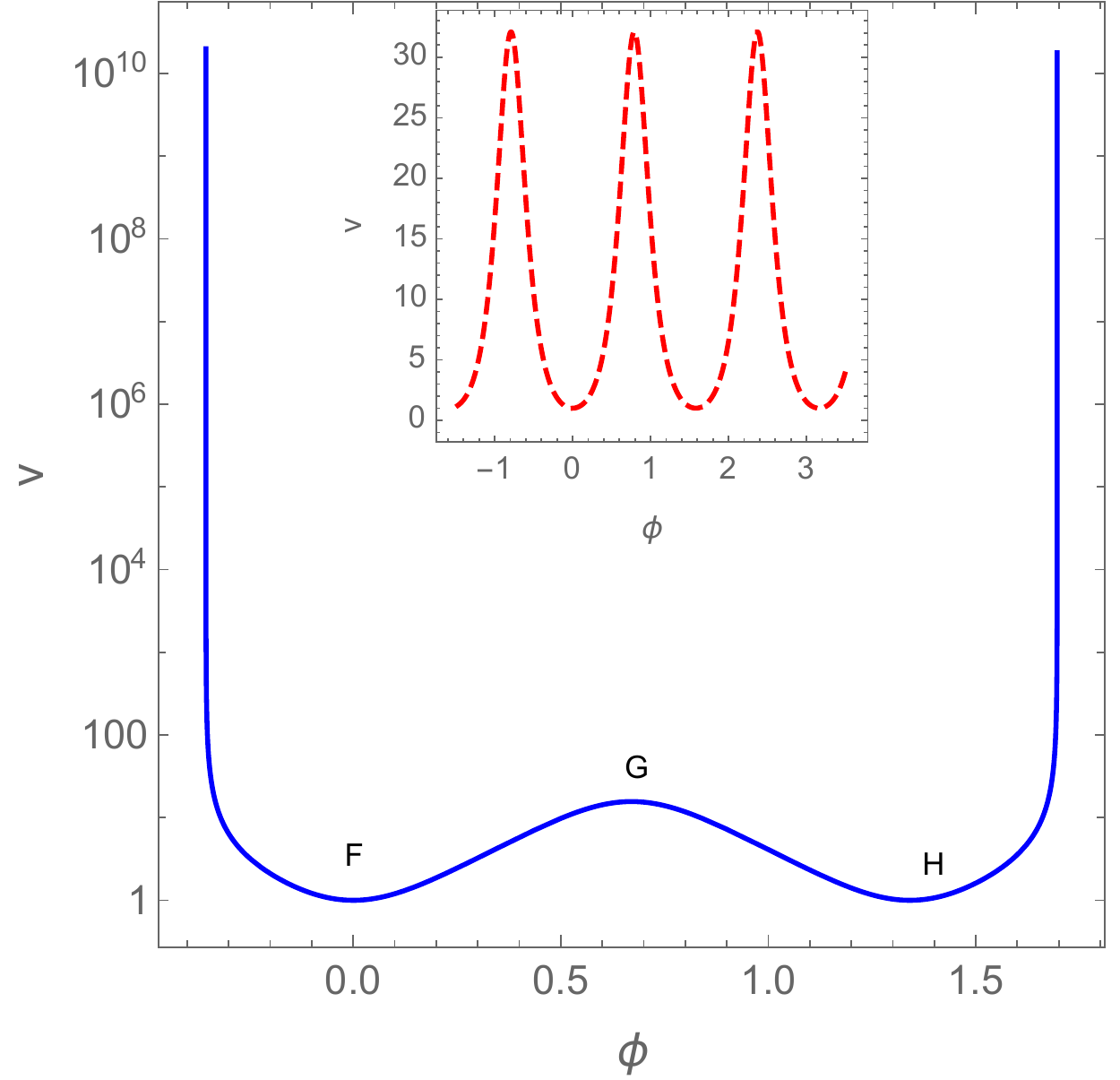}
\includegraphics[width=7cm]{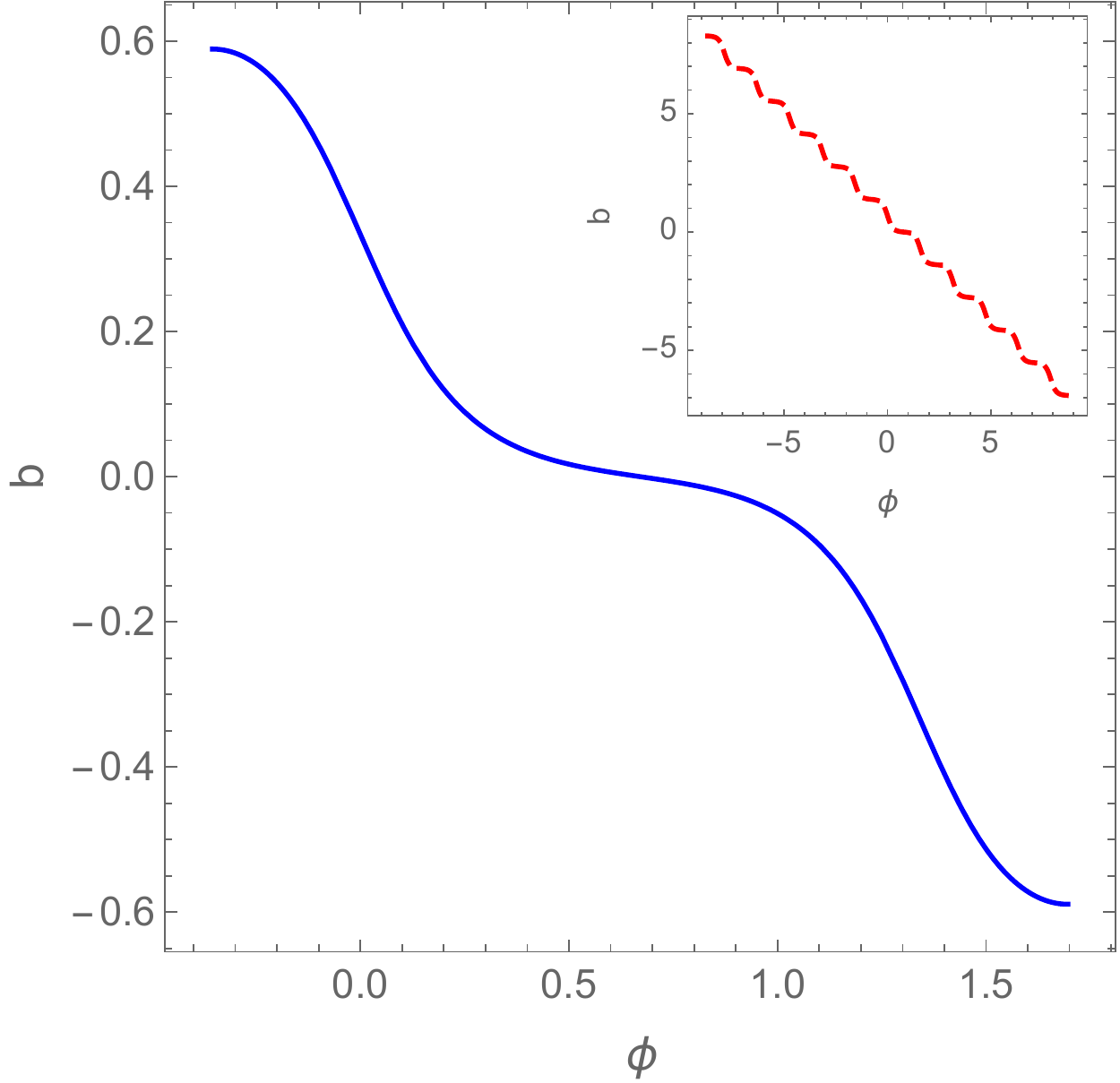}
\caption{Non-cyclic evolution for mLQC-I is shown and compared with the situation in LQC for initial conditions given in (\ref{initial1}). The blue solid curves depict the evolution of the volume and $b$ in mLQC-I with respect to the massless scalar field with the initial bounce located at $\phi=0$. The red dashed curves in the inset plots show the results in LQC with the same value of the cosmological constant  as in mLQC-I. The universe expands (contracts) rapidly in the distant future (past) in mLQC-I. In contrast, there appears a cyclic universe in LQC due to the effects of constant negative potential. }
\label{f2}
\end{figure}

\begin{figure} [tbh!]
\includegraphics[width=7cm]{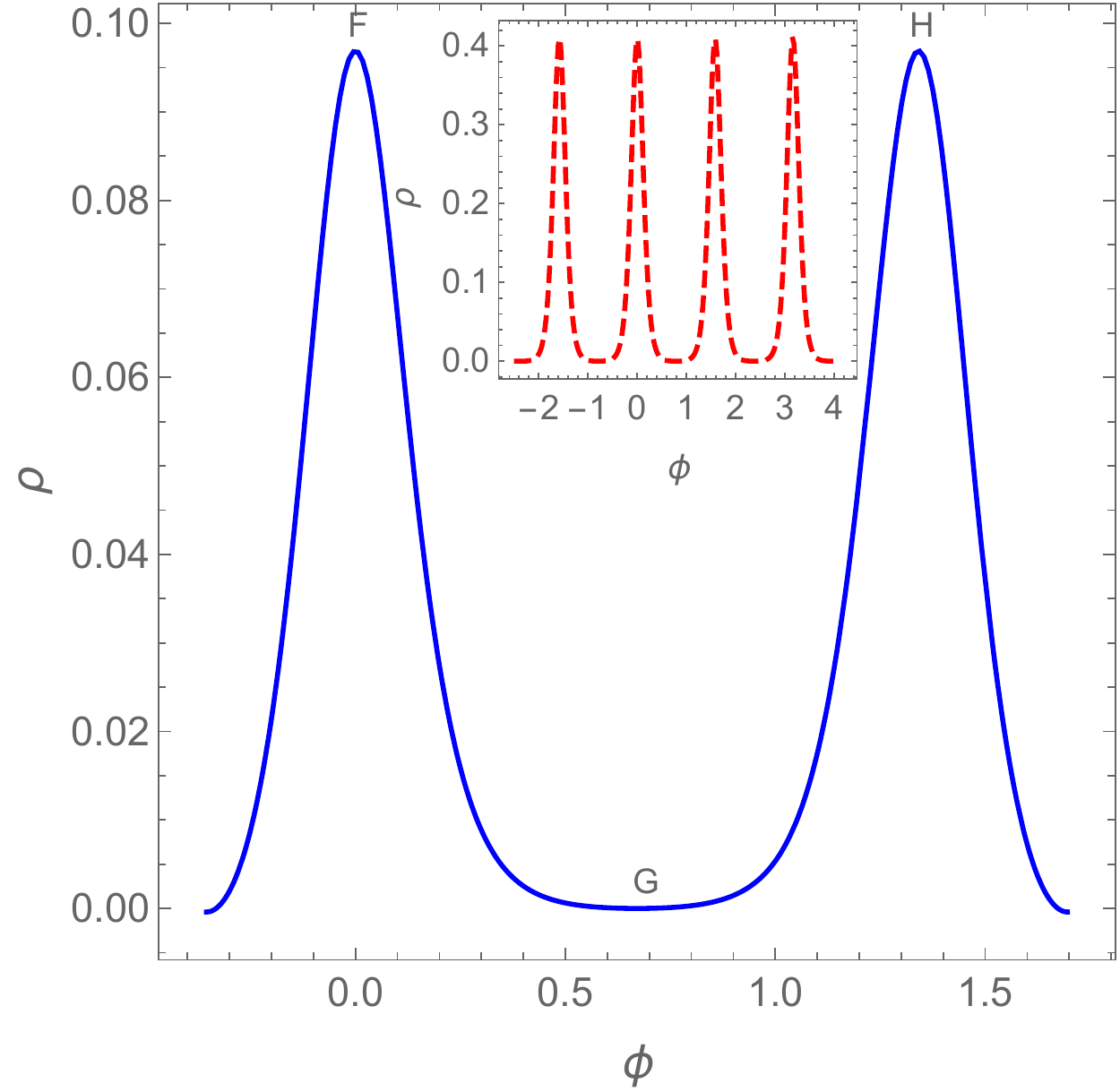}
\includegraphics[width=7cm]{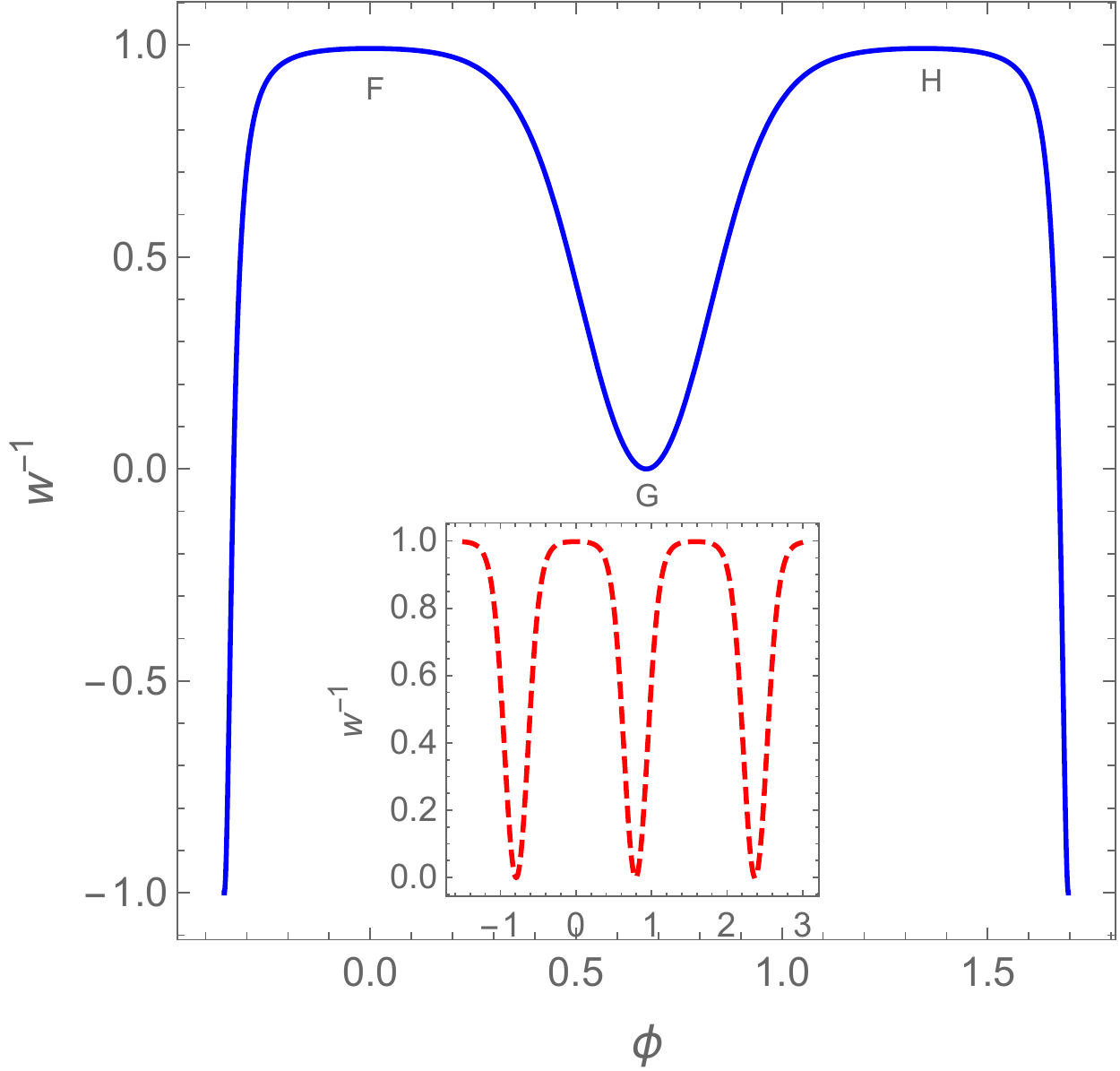}
\caption{In the figure, with the same initial conditions and the cosmological constant given in Fig. \ref{f2}, we show the evolution of the energy density and the inverse of the equation of state in mLQC-I (blue solid curves) and LQC (red dashed curves in the inset plots). In LQC, the energy density oscillates between $0$ and $\rho^\mathrm{LQC}_\mathrm{max}$ and the inverse of the equation of state lies in the range $w^{-1}\in(0,1)$. On the other hand, in mLQC-I, the energy density becomes negative in the distant past and future when the inverse of the equation of state lies in the interval $w^{-1}\in(0,-1)$ which implies a negative equation of state less than $-1$ and consequently the violation of the weak energy condition.} 
\label{eos for small cosmological constant}
\end{figure}

The numerical results for above initial condition are presented  in Figs. \ref{f2}-\ref{eos for small cosmological constant} where the blue solid curves are used for mLQC-I and the red dashed curves in the inset plots are for LQC. Fig. \ref{f2} depicts the behavior of $v$ and $b$ in both models. In the left panel (the volume plot), one finds that there are a total of two bounces in mLQC-I. The first bounce which takes place at $\phi=0$ corresponds to the point $F$ in Fig. \ref{mLQC-I} while the second bounce at around $\phi=1.40$ corresponds to the point $H$ in Fig. \ref{mLQC-I}. Between two bounces, there is a recollapse point at $\phi\approx 0.70$ which corresponds to the point $G$ in Fig. \ref{mLQC-I}. This recollapse point is in the $b_-$ branch and is of the first type as discussed in Sec. II where the energy density vanishes.
After the second bounce point $H$, the universe lies in the $b_+$ branch and quickly enters into a state of super-exponential expansion with a negative energy density. As the volume increases, the total energy density is dominated by the potential energy of the scalar field. From the Friedmann equation (\ref{b+}) and the relation (\ref{2.2}), one can find  when $\rho \rightarrow U_i=-3.98\times10^{-4}$,  $b\rightarrow -0.59$. Similarly, before the first bounce point $F$, the universe also lies in the $b_+$ branch and contracts super-exponentially in the distant past of this contracting phase when $\rho \rightarrow U_i$ and  $b\rightarrow 0.59$. As a result, during the whole evolution of the universe, $b$ is bounded by $b\in(-0.59,0.59)$ as depicted in the right panel (the $b$ plot) in Fig. \ref{f2} which also shows that $b$ monotonically decreases in the forward evolution (in the direction of increasing $\phi$). From the plots of volume and its conjugate momentum $b$, we see that there is no cyclic evolution in mLQC-I even though the potential energy is negative. This  is in a striking contrast with the LQC results depicted in the inset plots of Fig. \ref{f2} where the red dashed curve from LQC shows a cyclic universe in which all the recollapses happen at the zero energy density and all the bounces occur at the maximum energy density in LQC.  The volume of the universe at consecutive bounces or recollapses are exactly the same, implying that the universe would undergo identical cycles for an infinitely long time. Meanwhile, from the inset plot in the right panel of Fig. \ref{f2}, one can find the momentum $b$  changes monotonically in LQC and is not bounded from above or below.

In Fig. \ref{eos for small cosmological constant}, we show explicitly the behavior of the energy density and the inverse of the equation of state when the universe undergoes evolution as depicted in Fig. \ref{f2}. For a constant negative potential $U_i$, the pressure is always positive. Since at the recollapse point $G$ the equation of state would become infinite, in Fig. \ref{eos for small cosmological constant}, we show the plot of  $w^{-1}$ which takes value in a finite range in both LQC and mLQC-I. In LQC (red dashed curves in the inset plot), with the energy density oscillating between $0$ and $\rho_\mathrm{max}^\mathrm{LQC}$, the inverse of the equation of state changes correspondingly between $0$ and approximately $0.99$. This implies that the equation of state reaches its minimum value (larger than unity) at the bounce points and blows up at the recollapse points. Besides, there is no region in which the energy density would fall below zero. On the other hand, in mLQC-I (blue solid curves), the right panel of the Fig. \ref{eos for small cosmological constant} shows that the inverse of the equation of state lies in the range $w^{-1}\in(-1,1)$. In particular, $w^{-1}$ quickly becomes negative after (before) the bounce point $H$ ($F$) and moves towards $-1$. This implies that in mLQC-I, in the distant past and future, the energy density would become negative which results in a negative equation of state whose magnitude is always larger than unity. As a result, even though there is no cyclic evolution in mLQC-I, the weak energy condition is still violated in the distant past and future when the universe is filled with a constant negative potential (given by (\ref{initial1})).

\begin{figure}
\includegraphics[width=7cm]{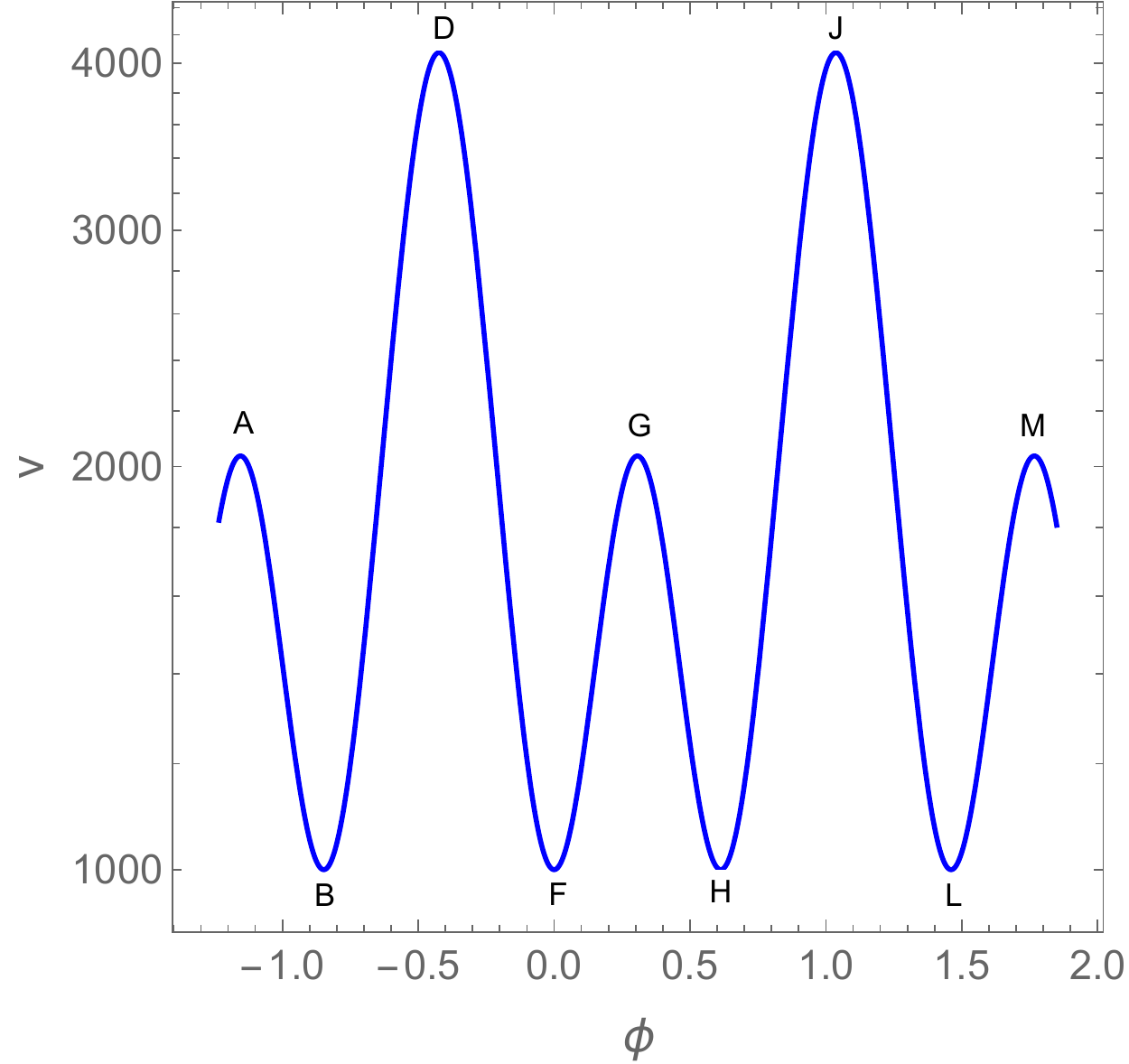}
\includegraphics[width=7cm]{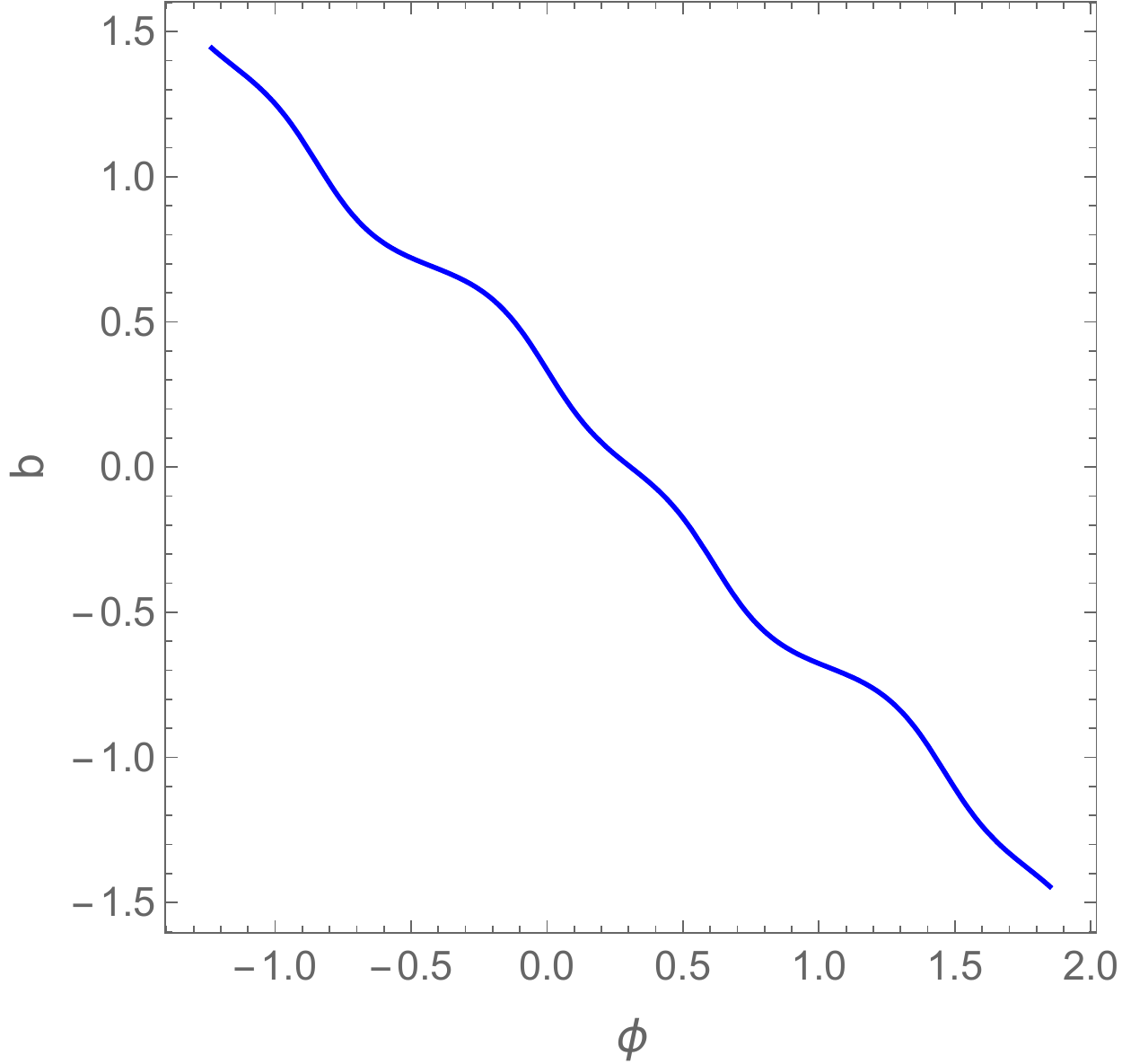}
\caption{The figure shows the case of Planckian sized negative  cosmological constant which allows a cyclic evolution. The evolution of the volume and $b$ in mLQC-I is depicted. The universe undergoes two distinct branches alternately, causing two local maxima of the volume in each cycle (for example point $D$ and point $G$). These maxima (and also minima) remain the same in successive cycles due to the constant potential. The momentum $b$ decreases monotonically and becomes unbounded. The bounce and recollapse  points in the $v$ plot  are labelled with the same letters as in Fig. \ref{mLQC-I}.}
\label{f3}
\end{figure}

\subsubsection{Cyclic evolution in mLQC-I with a Planckian negative cosmological constant}

\begin{figure}
\includegraphics[width=7cm]{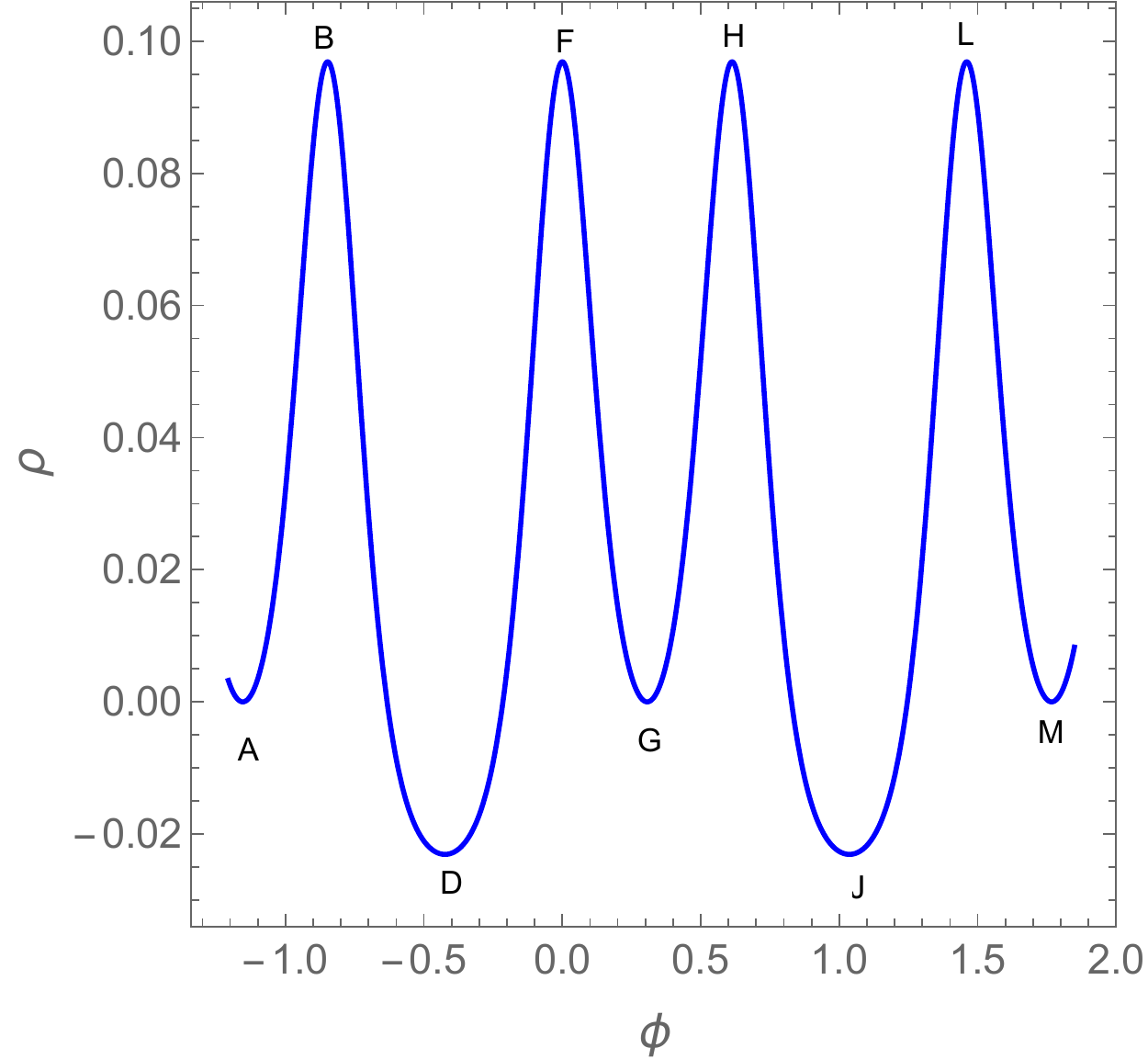}
\includegraphics[width=7cm]{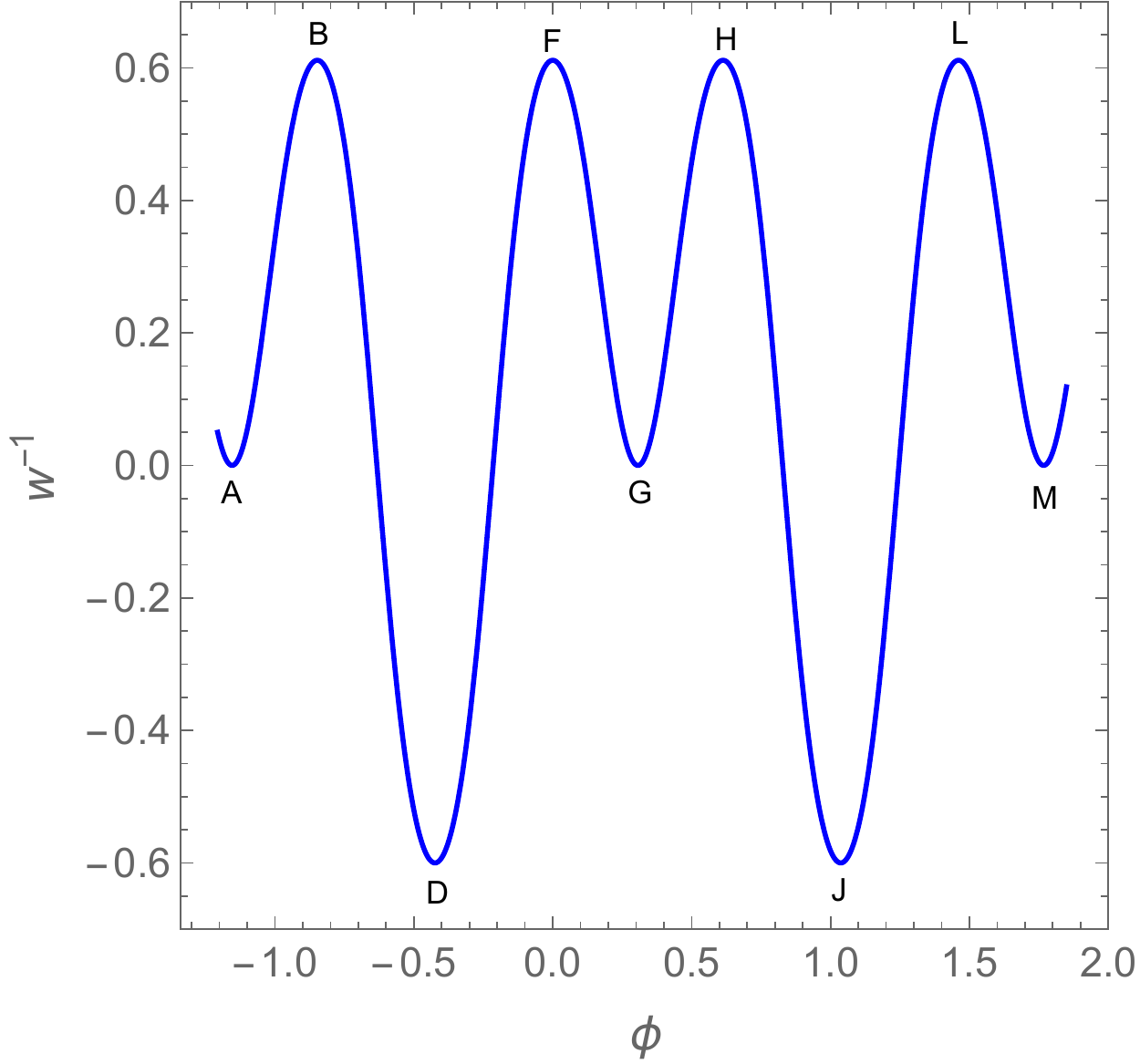}
\caption{The evolution of the energy density and the inverse of the equation of state in mLQC-I is shown for the same cosmological constant chosen in Fig. \ref{f3}, the energy density oscillates between $\rho_\mathrm{min}$ and $\rho_\mathrm{max}$ and the inverse of the equation of state also lies in a finite range which is about $|w^{-1}|<0.6$.}
\label{eos for large cosmological constant}
\end{figure}

As noted earlier, for a cyclic evolution in mLQC-I one needs to consider energy densities of the matter content which are more negative than $\rho_\mathrm{min}\approx-0.023$ (in Planck units). For such an example, we choose the cosmological constant to be
\bq
\label{initial2}
\Lambda=-\frac{4}{\lambda^2},
\eq
yielding $U_i=-0.03<\rho_\mathrm{min}\approx-0.023$ (in Planck units). The initial volume in this case is set to $1000$. The numerical results are presented in Figs. \ref{f3}-\ref{eos for large cosmological constant}. Since in this case the qualitative evolution of the universe in LQC remains the same as discussed in the previous case,  we only show the numerical solutions of mLQC-I in the figures. In LQC, the universe still evolves in identical cycles of contraction and expansion as depicted in the inset plots of Figs. \ref{f2}-\ref{eos for small cosmological constant}. There are only quantitative changes in LQC dynamics for different initial conditions, for example when $U_i$ decreases, the maximum volume of the universe would decrease and the minimal value of the equation of state would increase. On the other hand, there is a qualitative change in dynamics for mLQC-I where a cyclic universe appears as shown in Fig. \ref{f3}. While as in LQC, the momentum $b$ now becomes unbounded and evolves monotonically for a cyclic evolution of the universe, the characteristic feature of the cyclic universe in mLQC-I is that there exist two types of the recollapse points which yield two different recollapse  volumes of the universe. The first type of the recollapse points, such as points $A$, $G$ and $M$ belong to the $b_-$ branch, while the second type of the recollapse points, such as $D$ and $J$, belong to the $b_+$ branch. As shown in the  $\rho$ plot of Fig. \ref{eos for large cosmological constant}, these two types of recollapse points also have different recollapse energy densities. To be specific, the energy density becomes zero at the recollapse points in the $b_-$ branch on one hand and reaches $\rho_\mathrm{min}$ at the recollapse points in the $b_+$ branch on the other hand. It is important to note that though the universe is cyclic in this case, it retains its Planckian character for a large part of  evolution as is evident from the behavior of energy density.

Since the equation of state becomes infinite at each recollapse point in the $b_-$ branch, we show the $w^{-1}$ plot in the right panel of Fig. \ref{eos for large cosmological constant} which indicates the magnitude of $w^{-1}$ is confined within the range $w^{-1}\in(-0.6,0.6)$. As a result,  the magnitude of the equation of state is always larger than one. More specifically,  in the neighborhood of the recollapse and bounce points in the $b_-$ branch, the equation of state is positive and attains its minimum at each bounce point. Whereas, in the neighborhood of the recollapse points in the $b_+$ branch, $w^{-1}$ becomes negative implying that the equation of state becomes less than $-1$. Therefore, the weak energy condition is violated in the neighborhood of the recollapse points (such as points $D$ and $J$) in the $b_+$ branch.
It should be noted that for other negative cosmological constants with larger magnitudes, although the exact values of the local maxima of the volume  would change correspondingly, the qualitative dynamics of the evolution of the universe in mLQC-I remains  the same as  depicted  in Figs. \ref{f3}-\ref{eos for large cosmological constant}.

To summarize the discussion for the negative cosmological constant, our numerical results confirm that unless the matter content is such that a Planckian sized negative cosmological constant is present there is no cyclic evolution in mLQC-I. Further there is a violation of weak energy condition   near the points of recollapse. Notably the cyclic universe obtained is highly quantum and Planck sized as is evident from the figures of volume and energy density \ref{f3}-\ref{eos for large cosmological constant}. This observation is in contrast to LQC where a macroscopic cyclic universe can be generically obtained for a constant negative potential without any violation of weak energy condition.

\subsection{The effective dynamics with a cyclic potential}

In this subsection,  we discuss the background dynamics of  the  universe in mLQC-I and LQC in presence of a negative potential. The cyclic potential used in  the following originates from  an ansatz for a cyclic universe proposed by Steinhardt and Turok \cite{steinhardt2002}. It was later studied in \cite{lqc-cyclic1,lqc-cyclic2,lqc-cyclic3} within the framework of LQC.
The form of the cyclic potential is given by
\bq
\lb{3b1}
U=\alpha \left(1-e^{- \sigma  \phi}\right)\mathrm{exp}\left(-e^{-\omega \phi}\right),
\eq
where parameter $\alpha$ is chosen to give the right magnitude of the current cosmic acceleration and other parameters can be fixed by the current observations on the density perturbations.  The potential $U$ tends to  $\alpha$ ($>0$) when $\phi\rightarrow \infty$, and decreases monotonically as $\phi$ moves towards the origin.  It attains its minimum  negative  value at some point in the regime $\phi<0$ and goes to zero as $\phi\rightarrow -\infty$. In the Ekpyrotic model, the potential is always positive (negative) when $\phi>0$ ($\phi<0$) while the location and the value of the minimum of the potential are determined by the parameters $\alpha$, $\sigma$ and $\omega$ .  Without any loss of generality, in order to study the qualitative dynamics resulting from the potential, one can fix the parameters to $\sigma=0.3\sqrt{8\pi}$, $\omega=0.09\sqrt{8 \pi}$ as in \cite{lqc-cyclic2} while choosing different values of $\alpha$ to investigate the effects of the minimum of the potential on the background evolution  of the  universes in  LQC and mLQC-I.  In the following, we discuss two representative choices of $\alpha$  which yield qualitatively distinct dynamics as the minimum of the potential changes.

Before going into the details of the numerical solutions, we would like to emphasize that  the initial conditions in this case  are chosen in such a  way that the scalar field is initially released from the right wing of the potential  and moving towards the bottom of the potential when the universe is initially in a state of contraction. The bounce occurs later when the potential energy become negative. As a result, one  starts with a small positive value of $\phi$ and a negative momentum of the scalar field. In this way, the initial energy density is fixed and $b$ can be solved from the effective Hamiltonian constraint. Similar to the cases in the previous subsection, we also choose a positive $b$ which is closest to zero.

\subsubsection{Non-cyclic evolution in mLQC-I}

\begin{figure}
\includegraphics[width=8cm]{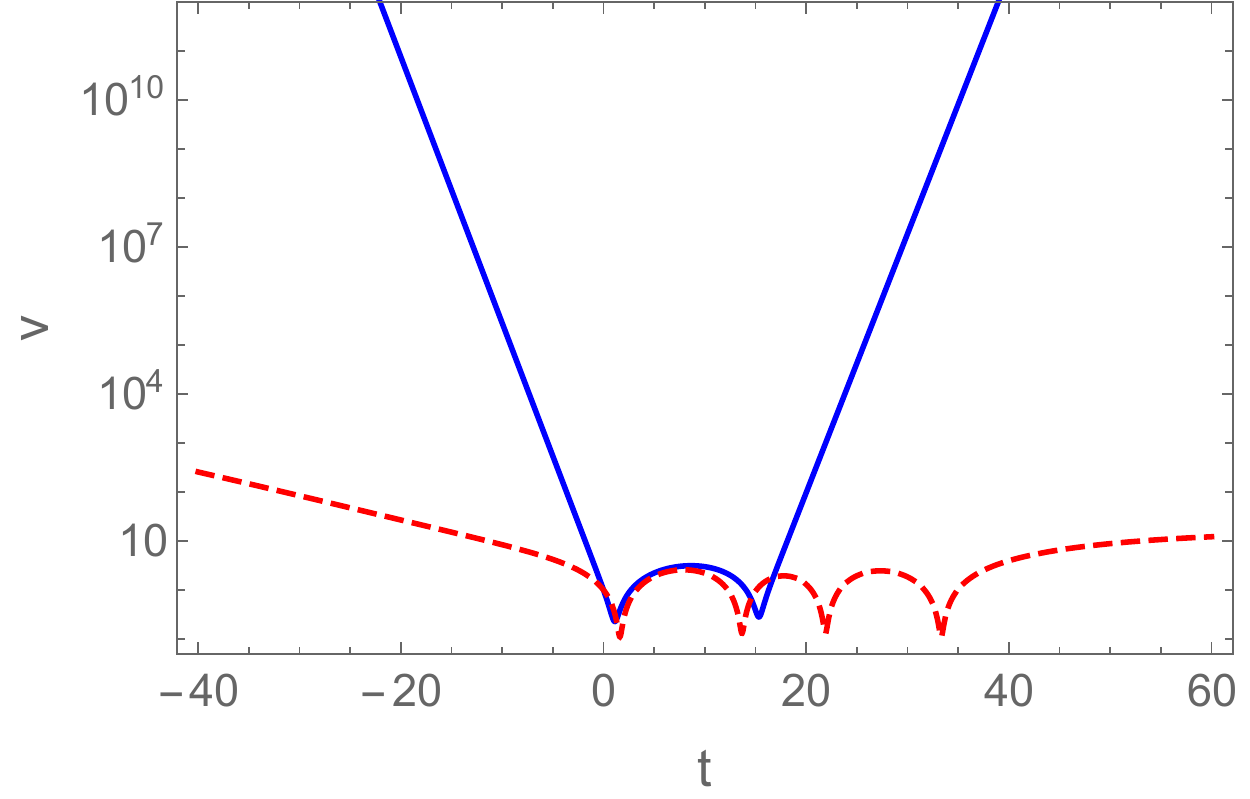}
\includegraphics[width=8cm]{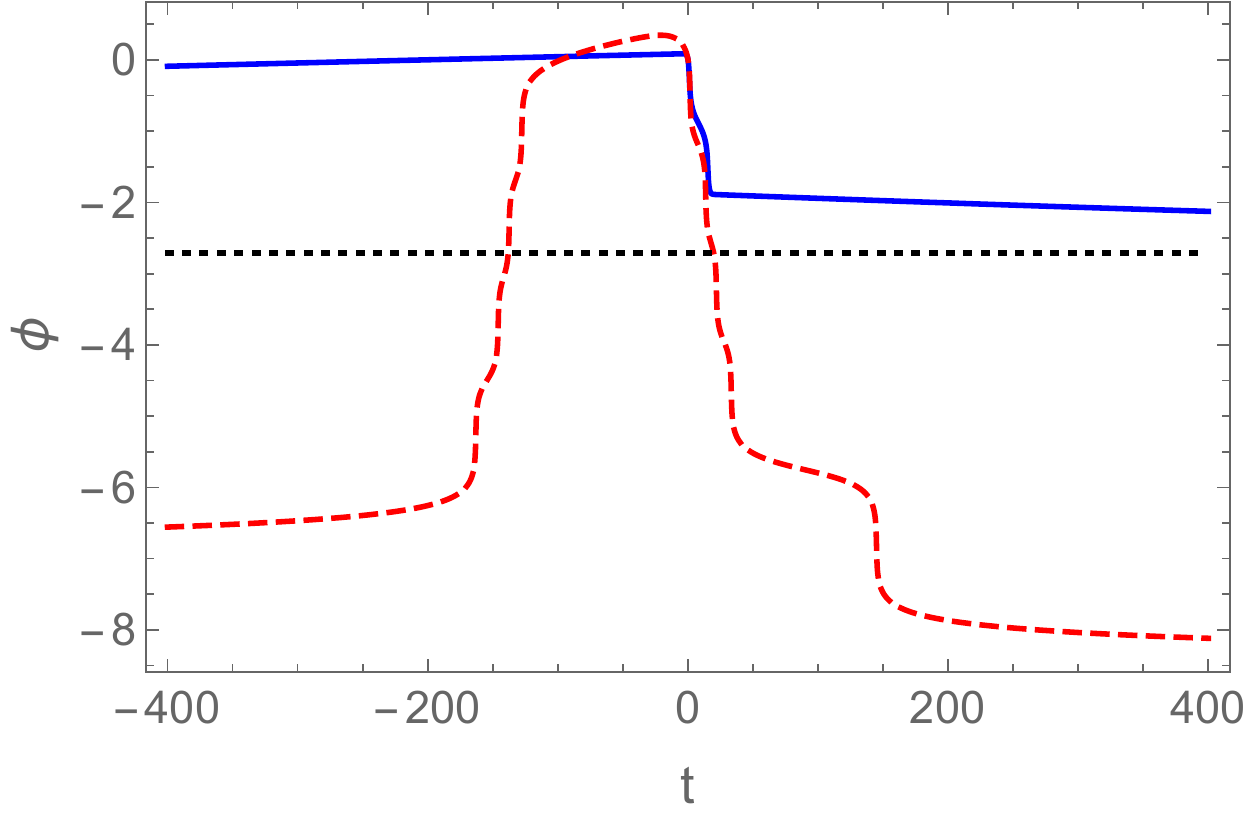}
\caption{These plots show non-cyclic evolution for the cyclic potential in mLQC-I and compare to the cyclic evolution in LQC. With $\alpha$ and the initial conditions  given in (\ref{initial3}), the blue solid curves depict the evolution of the volume (left panel) and the scalar filed (right panel) in mLQC-I while the red dashed curves are the results from LQC. The  black  dotted line in the right panel shows the value of the scalar field at which the minimum of the  potential is attained. }
\label{f4}
\end{figure}

\begin{figure}
\includegraphics[width=8cm]{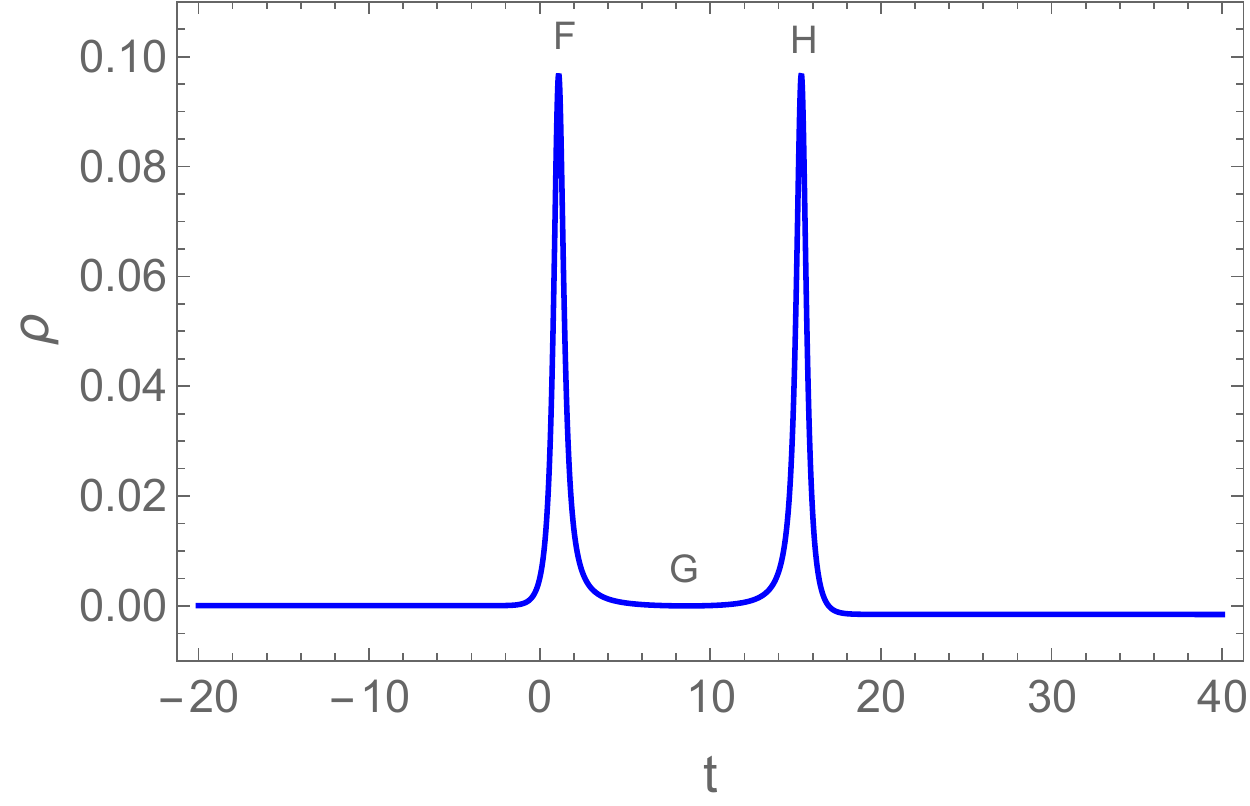}
\includegraphics[width=8cm]{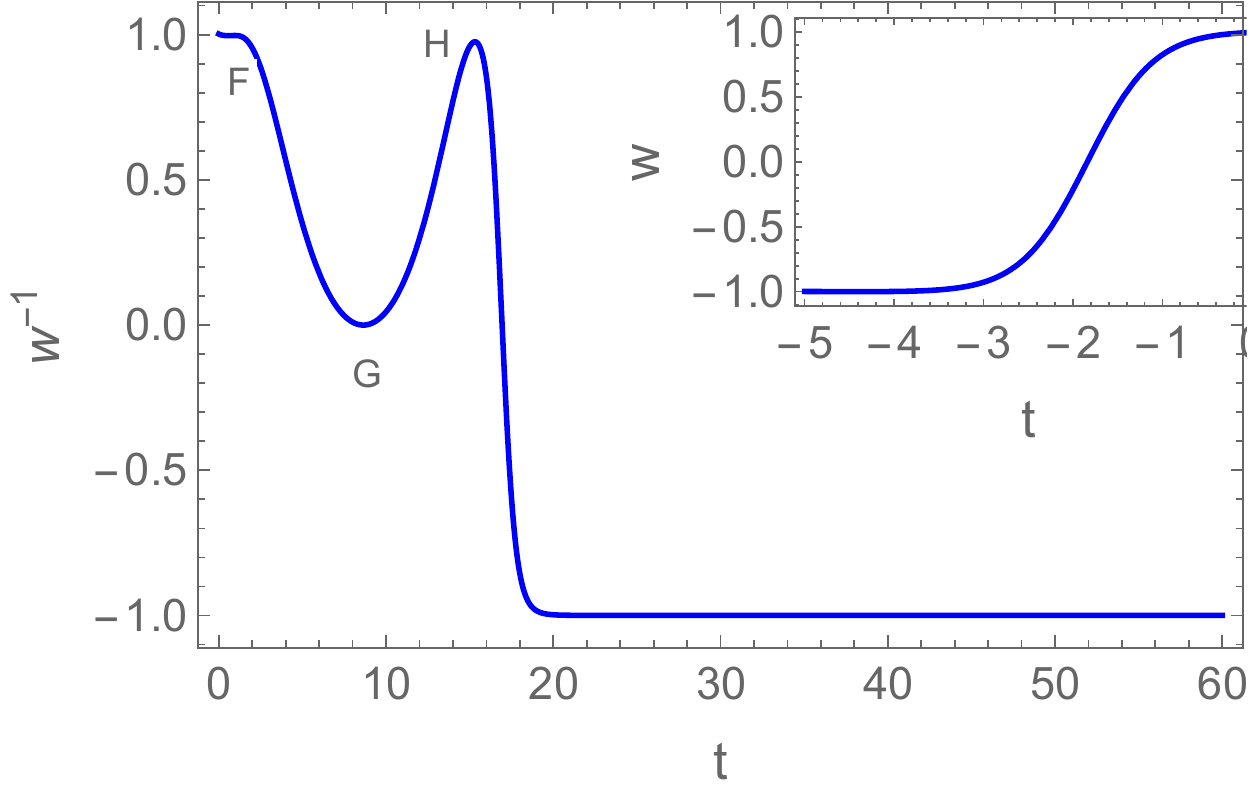}
\caption{In this figure, we show the energy density and the inverse of the equation of state in mLQC-I with the same $\alpha$ and the initial conditions as in Fig. \ref{f4}. Since the minimal value of the scalar potential is larger than $\rho_\mathrm{min}$ in mLQC-I, there only exist two bounces ($F$ and $H$) as shown in the left panel. In the right panel, we show the inverse of the equation of state in the time range $t\in(0,60)$ where $w^{-1}$ vanishes at the recollapse point $G$ and then quickly tends to negative unity. The inset plot shows the  behavior of the equation of state in the time range $t\in(0,-5)$ where the equation of state tends to negative unity corresponding to an emergent de Sitter phase.}
\label{energy density for small scalar potential}
\end{figure}

In the first example,  the parameter $\alpha$, the initial value of the scalar field  and its momentum are set to
\bq
\lb{initial3}
\alpha=0.001, \quad \quad  \phi_i=0.01,\quad \quad v_i=1,\quad \quad p_{\phi_i}=-0.10,
\eq
so that the initial energy density $\rho_i=5.01\times10^{-3}$, the initial value of $b$ is $0.58$ in mLQC-I and $1.33$ in LQC. With $\alpha=0.001$,  the potential  reaches its minimum value, namely $U_\mathrm{min}=-0.0019$ at $\phi=-2.7$. Note in this case $U_\mathrm{min}>\rho_\mathrm{min}$.  The numerical results are presented in Figs. \ref{f4}-\ref{energy density for small scalar potential} where the blue solid curves are used for mLQC-I and the red dashed curves are used for LQC. Since the scalar field becomes massive in this case,  it cannot play the role of a good global clock any more, and we use cosmic time to study evolution.

Similar to the  case in Fig. \ref{f2},  the universe in mLQC-I undergoes a total of two bounces at points $F$ and $H$ as shown in the left panels of Figs. \ref{f4}-\ref{energy density for small scalar potential}. However, due to the asymmetry of the potential, the evolution is now asymmetrical with respect to the recollapse point $G$. Since the energy density can never decrease to  $\rho_\mathrm{min}$ in the current case, the universe can never arrive at the recollapse points $D$ and $J$ in the $b_+$ branch. As a result, after the second bounce at point $H$, the universe  quickly enters into a state of super-exponential expansion with a Planck-scale Hubble rate and the energy density decreases monotonically ever after.  Meanwhile, in the right panel of Fig. \ref{f4},  we can see in mLQC-I the scalar field rolls down the potential and moves towards the bottom of the potential, which is represented by the black dotted line in the plot.  This is in contrast with a case in LQC plotted in the red dashed curves. The universe in LQC undergoes a number of bounces in a short period of time between $t=0$ and $t=40$ when the scalar field passes through the bottom of the potential. Then, the universe enters into an expanding phase when the scalar field moves away from  the potential well and rolls on the left wing of the potential. The maxima of the volume at each  consecutive recollapse change in small amounts due to non-constant potential.

In Fig. \ref{energy density for small scalar potential}, we show the behavior of the energy density and the inverse of the equation of state in mLQC-I. In LQC, the energy density is always non-negative and there is no violation of the weak energy condition at any stages of the evolution, so we only show the plots for mLQC-I for brevity. In the right panel of the figure, we plot the inverse of the equation of state for $t\in(0,60)$ which shows two maxima at the bounce points $F$ and $H$. After the second bounce at $H$, the inverse of the equation of state quickly goes below zero and moves towards negative unity, implying an equation of state less than $-1$ and thus the violation of the weak energy condition. In the contracting regime before the bounce point $F$, we find an equation of state which ranges in the interval $w\in(-1,1)$ as shown in the inset plot of Fig. \ref{energy density for small scalar potential}. In this case, before the bounce at point $F$ there appears a brief quasi de Sitter phase, indicating a small positive energy density. This is because the scalar field is moving on the right wing of the potential and thus the potential energy is positive in this regime. We find the equation of state turns out to be less than $-1$ again if the universe is evolved backwards to any time at $t<-205.5$ (at $t=-205.5$, the energy density vanishes which corresponds to the point $E$ in Fig. \ref{mLQC-I}), which implies the energy density becomes negative and the weak energy condition is violated in the distant past when $t<-205.5$.

\begin{figure}
\includegraphics[width=8cm]{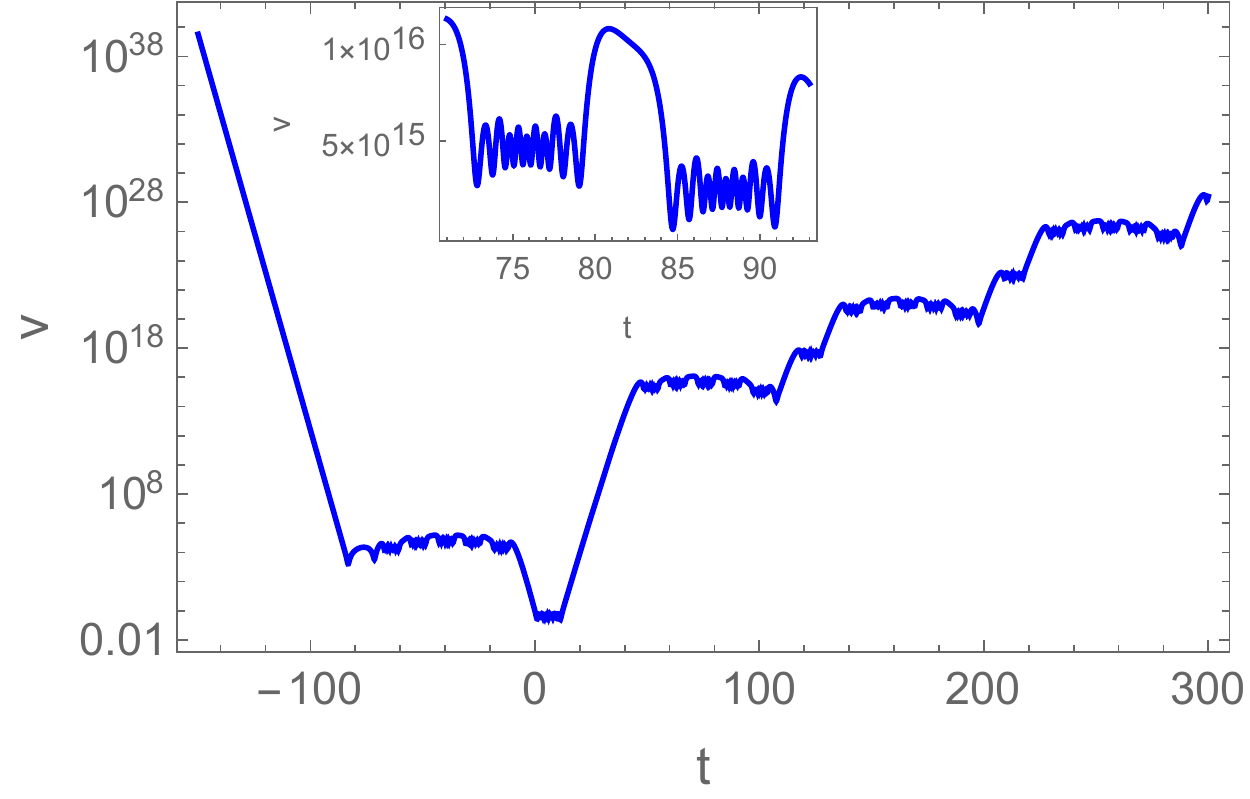}
\includegraphics[width=8.3cm]{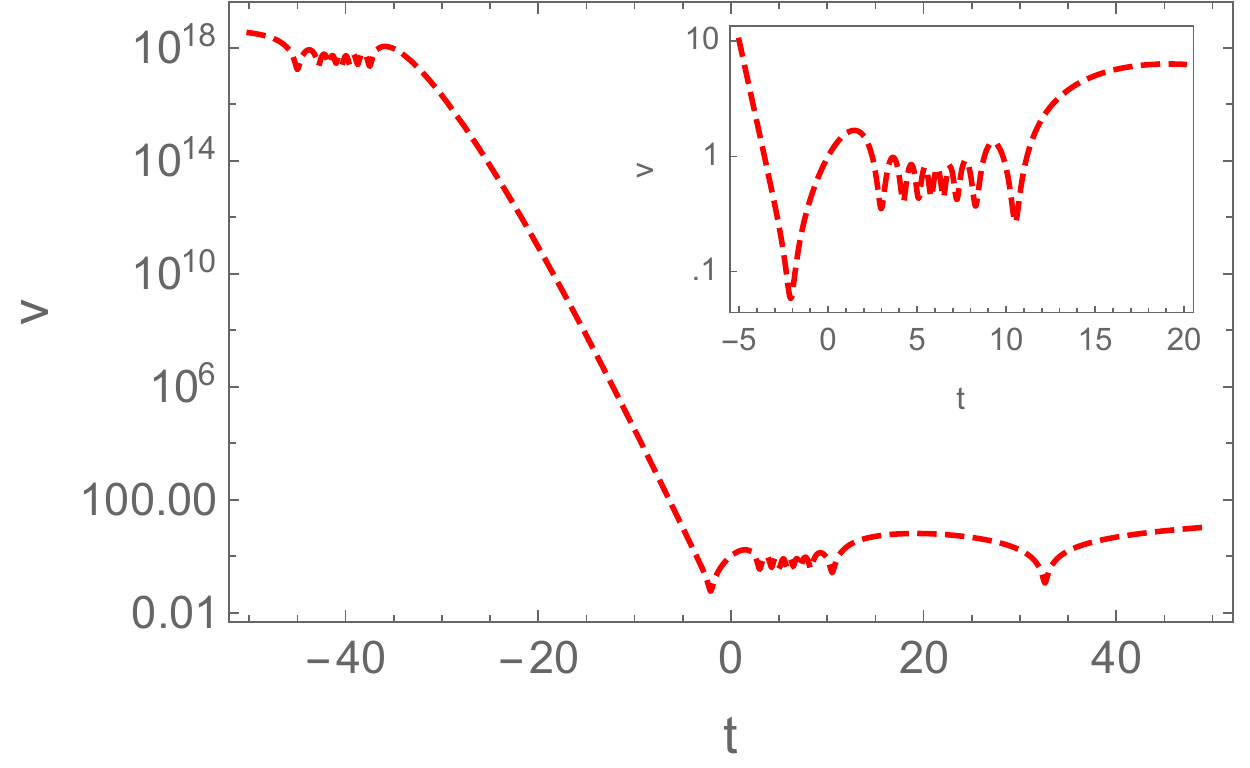}
\caption{With initial conditions given in (\ref{initial4}), the evolution of the volume is displayed for mLQC-I (left panel) and LQC (right panel). Islands of cluster of bounces separated by rapid expansions or contractions can be observed in mLQC-I and LQC. The inset plots zoom in around one cluster and show explicitly the bounces and the change in the maximum volumes at consecutive recollapse points.}
\label{f5}
\end{figure}

\begin{figure}
\includegraphics[width=8cm]{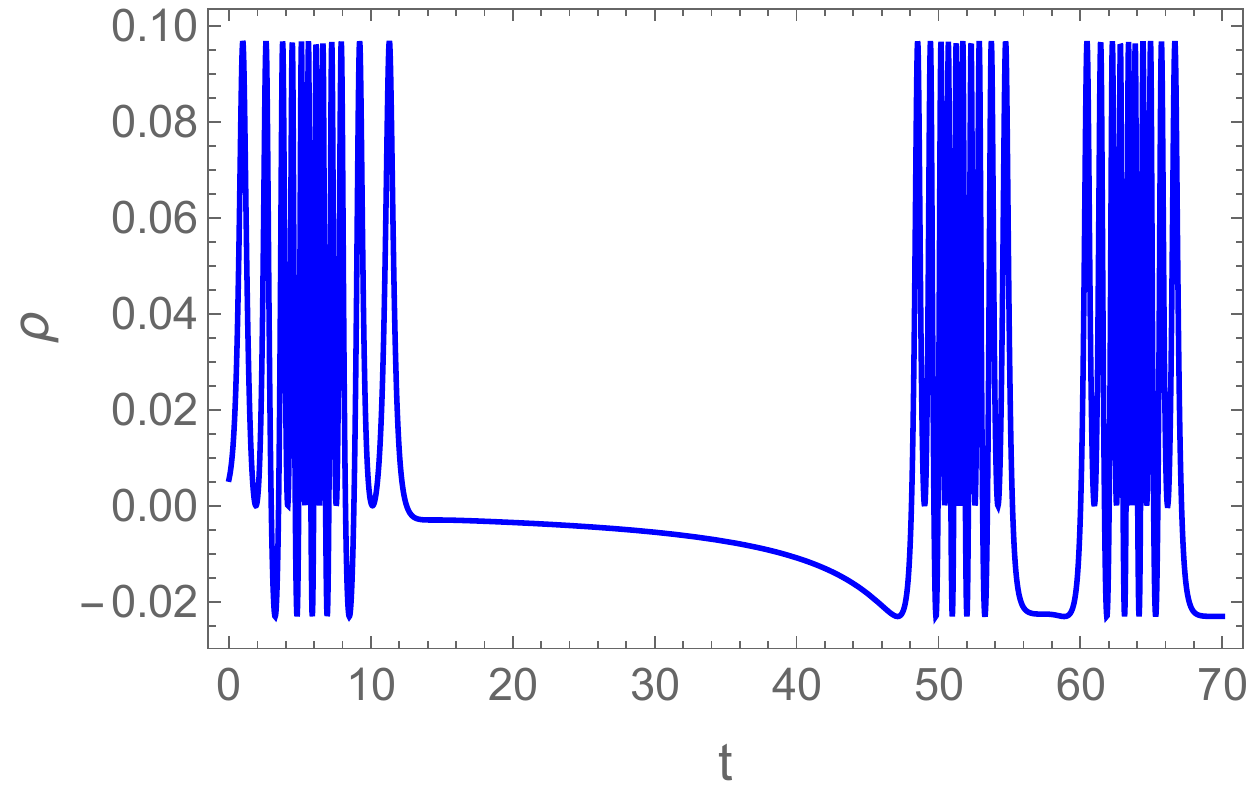}
\includegraphics[width=8cm]{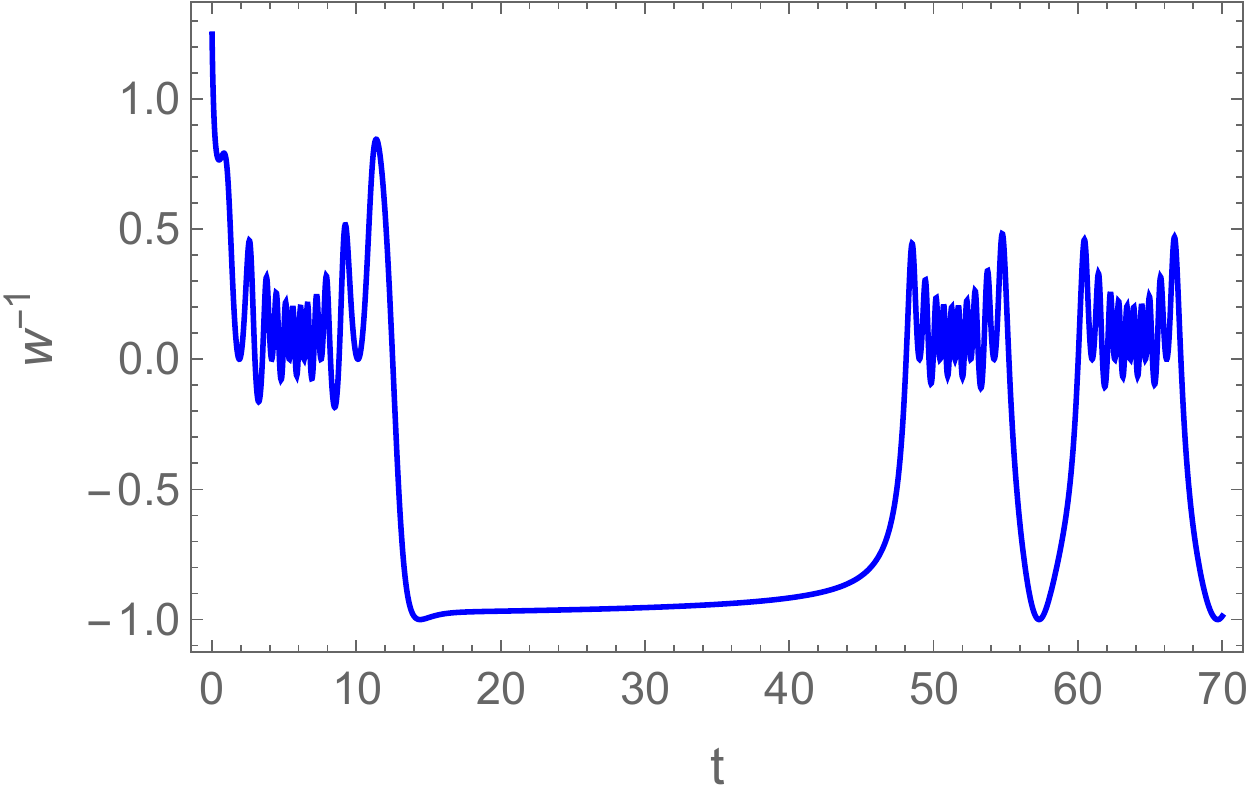}
\caption{With the same initial conditions as in Fig. \ref{f5}, we show the behavior of the energy density and the inverse of the equation of state in the time range $t\in(0,70)$, where both of them oscillate rapidly in each cluster of bounces. In each cluster and also between two neighboring clusters, there are times when the energy density turns out to be negative and the equation of state becomes less than $-1$ which are indicative of the violations of the weak energy condition.}
\label{energy density for large scalar potential}
\end{figure}

\subsubsection{Cyclic evolution in mLQC-I with a Planckian negative potential}
We now consider the case where negative potential has an almost Planckian magnitude. This
 example is depicted in Figs. \ref{f5}-\ref{energy density for large scalar potential} which display the numerical results with the initial conditions
\bq
\lb{initial4}
\alpha=0.10, \quad \quad  \phi_i=0.01,\quad \quad v_i=1,\quad \quad p_{\phi_i}=-0.10,
\eq
so that the initial energy density $\rho_i=5.55\times10^{-3}$, the initial value of $b$ is $0.58$ in mLQC-I and $0.05$ in LQC. In this case, the bottom of the potential  is located at $\phi=-2.7$ with $U_\mathrm{min}=-0.19<\rho_\mathrm{min}$.  In spite of the change in $U_\mathrm{min}$, the qualitative dynamics of the universe in LQC (depicted in the right panel of Fig. \ref{f5}) is  similar to Fig. \ref{f4}. The  difference lies in the fact that since the potential well becomes deeper in this case, there are more bounces located very close to one another (as shown in the inset plot of the right panel in Fig. \ref{f5}). These clusters of bounces are produced as the scalar field moves across the bottom of the potential. Starting from $t=-50$, the scalar field is passing through the potential well from the left, resulting in the first cluster of bounces at $t\approx-40$. Then, it reaches the turnaround point on the right and rolls back to the left, causing the second cluster of bounces near $t=5$.  During this process,  the scalar field does not oscillate at the bottom of the potential  in LQC so there are only two clusters of bounces as the scalar field traverses one complete cycle on the potential.  The situation becomes more complicated in mLQC-I as shown in the left panel of Fig. \ref{f5} where a number of islands of clusters of bounces separated by super-exponential expanding or contracting phases can be observed. These islands are formed when the scalar field keeps oscillating at the bottom of the potential. The inset plot of the left panel of Fig. \ref{f5} displays two clusters of bounces in mLQC-I in which the maximum volumes at consecutive collapses do not uniformly increase or decrease. Here we note that such islands of clusters of bounces have been found numerically earlier also in LQC but in the presence of spatial curvature  \cite{hysteresis2,hysteresis3}.  We plot the energy density and the inverse of the equation of state in mLQC-I in Fig. \ref{energy density for large scalar potential}. From the left panel, one can find there are regimes in each cluster of bounces where the energy density goes to negative and reaches $\rho_\mathrm{min}$. Correspondingly, the $w^{-1}$ lies between zero and $-1$ in the regimes when the energy density is negative, implying an equation of state less than negative unity and thus the violation of the weak energy condition.

To end this section, we conclude that our numerical results are consistent with the analysis in Sec. \ref{sec:analytical analysis}. In LQC, as generally in bouncing models,  a necessary condition to produce a cyclic evolution in a universe is to introduce a negative potential which can lead to recollapses  with vanishing energy density. During the whole evolution, the energy density in LQC is always non-negative. However, a negative potential can not guarantee a cyclic universe in mLQC-I. A cyclic universe is only allowed in mLQC-I if one chooses an almost Planckian valued negative potential. Moreover, the cyclic evolution of the universe in mLQC-I can only be realized when the total energy density decreases to  $\rho_\mathrm{min}\approx-0.023$ at the recollapse points in the $b_+$ branch, that is to say, the weak energy condition must be violated to allow a cyclic evolution in mLQC-I. Finally, even in the cases where a cyclic evolution is allowed in mLQC-I there is almost no classical epoch in the dynamics and the energy density oscillates rapidly between its maximum and minimum values.

\section{Summary}
\label{summary}
\renewcommand{\theequation}{4.\arabic{equation}}\setcounter{equation}{0}

Previous studies on loop quantum cosmologies were mainly focused on the investigation of the robustness of the singularity resolution in various isotropic and anisotropic spacetimes as well as for different regularizations of the Hamiltonian constraint. Since it was generally expected that a cyclic universe is guaranteed once the singularity is resolved and a negative potential is included, not much attention was so far paid to studying whether a cyclic evolution of the universe is also a generic feature of loop cosmologies. The goal of this paper was to demonstrate that quantum gravitational modifications responsible for singularity resolution may modify the dynamics in such a way that a cyclic evolution becomes difficult to achieve. While in the standard LQC, a spatially-flat universe can naturally undergo a cyclic evolution without violating any energy conditions  once a scalar field with a negative potential or a negative cosmological constant is introduced, we showed that in the LQC employing Thiemann's regularization of the Hamiltonian constraint, namely mLQC-I, a cyclic universe can be realized only under highly restrictive conditions which do not allow a classical regime.
For a large region of parameter space, the universe can  at most have only two bounces and a single recollapse in mLQC-I in contrast to infinite number of bounces and recollapses in LQC.

Assuming the validity of effective dynamics in LQC and mLQC-I, we first analyzed the generic features of the evolution of a spatially-flat FLRW universe with a constant and varying negative potential and found the main distinctions between two loop cosmological models. In standard LQC, since the contracting and the expanding phases are described by the same modified Friedmann equation, the bounces always take place at the maximum energy density and the recollapses at the minimal energy density which is zero in LQC. As a result, the minimal requirement to have a cyclic universe in LQC is the presence of a negative potential with no specifications on the magnitude of the minimal of the potential. Besides, at any stage, the energy density of the universe in LQC is always non-negative which implies that  the equation of state of the scalar field is always greater than unity. The nature of the recollapse is the same in each cycle which occurs in the classical regime. On the other hand,  in mLQC-I, there are two qualitatively distinct branches, namely the so-called $b_+$ and $b_-$ branches. These two branches are described by two different modified Friedmann equations and are connected via a quantum bounce where the energy density reaches its maximum value in the model. The $b_-$ branch can describe either a  contracting or an expanding phase which has similar properties to the contracting or the expanding phase in LQC in the sense that an expanding phase in the $b_-$ branch would switch into a contracting phase at a recollapse point where the energy density becomes zero. Moreover, in the $b_-$ branch, the energy density is always non-negative and the classical GR can be recovered when the energy density is far below the Planck scale. In contrast, the $b_+$ branch has very distinctive features. Although it can also account for both expanding  and  contracting phases, an expanding phase in the $b_+$ branch can only switch into a contracting phase at the recollapse point with a negative energy density $\rho_\mathrm{min}$ which is also the minimal energy density allowed in the model. As a result, the energy density in the $b_+$ branch must be negative with a magnitude of at least about $2\%$ of the Planck density near the recollapse point, resulting in a violation of the weak energy condition. Moreover, the GR limit can not be recovered in the $b_+$ branch due to the emergence of a Planckian magnitude effective cosmological constant at the vanishing energy density.  With the above properties of the two branches in mLQC-I in mind, we found a cyclic universe in mLQC-I must go through each branch  alternately as the quantum bounce would switch one branch into another automatically. The minimal requirement for the realization of a cyclic universe in mLQC-I is then  to reach the recollapse point in the $b_+$ branch, which implies that  the minimal of the potential of the scalar field must be more negative than  $\rho_\mathrm{min}$, considering the kinetic energy density of the scalar field is always non-negative. Since the $b_+$ branch does not exist in the classical GR but only emerges due to a separate regularization of the Lorentzian part of the classical Hamiltonian constraint, it should be regarded as one of the potential outcomes of the quantum geometrical effects.

For the negative cosmological constant, we tested two different magnitudes with one less negative than $\rho_\mathrm{min}$  and the other more negative than $\rho_\mathrm{min}$. We found a cyclic evolution of the universe in both cases in LQC while in mLQC-I, a cyclic universe can be observed only for the second case in which the cosmological constant is chosen to be more negative than $\rho_\mathrm{min}$. The evolution of the cyclic universe in LQC is characterized by the same energy density ($\rho=0$) and the same maximum volume at all of the recollapse points as well as an equation of state greater than unity. As a result, the weak energy condition is always satisfied in LQC.  On the other hand, in mLQC-I, when the cosmological constant is less negative than $\rho_\mathrm{min}$, only one recollapse point which belongs to the $b_-$ branch would appear, in the distant past and future, the universe would always remain in the $b_+$ branch since the energy density can not attain the value required by the recollapse point in the $b_+$ branch. It is only when  the energy density associated with the cosmological constant becomes more negative than $\rho_\mathrm{min}$ then a cyclic universe emerges in mLQC-I which is characterized by two alternating distinct recollapse points with different recollapse volumes.

The situation becomes much richer when it comes to a non-constant potential. In LQC, a cyclic universe can be observed regardless of the magnitude of the minimal value of the negative potential. However, due to the asymmetry of the potential, the qualitative behavior of the cyclic universe is no longer the same as the identical cycles observed in the case of a negative cosmological constant. There appear some regions with clusters of bounces where rapid contractions and expansions are located near to each other, these regions correspond to the moments when the scalar field is passing through the bottom of the potential, while in other regions when the scalar field is rolling on the wings of the potential,  exponential contraction or expansion is observed in a longer time period.  On the contrary, the cyclic evolution of the universe can be observed in mLQC-I only when the minimal of the potential is more negative than $\rho_\mathrm{min}$. We also observe more clusters of bounces in the case of mLQC-I which are formed when the scalar field is oscillating around the bottom of the potential.

In summary, the main result of our manuscript is that  the realization of a cyclic universe in mLQC-I is highly restricted as compared with in LQC. We wish to emphasize that this is due to the presence of the $b_+$ branch in mLQC-I which has no classical counterpart in the classical GR and thus should be regarded as an outcome of the regularization prescription when gravity is loop quantized. More specifically, the recollapse point in the $b_+$ branch can be reached only when the energy density becomes negative with an almost Planck-sized magnitude. As a result, near the recollapse point in the $b_+$ branch the weak energy condition is violated in mLQC-I which is in contrast with the cyclic universe in LQC where there is no violation of any energy conditions at any stages of evolution. Therefore, unlike the resolution of the curvature singularities which is a generic feature with respect to quantization ambiguities, the cyclic evolution of the universe can be highly restricted by the quantization prescriptions in loop quantum cosmologies. Finally, our study opens a new avenue to phenomenologically constrain loop quantum cosmological models. It shows that certain regularizations in LQG are simply incompatible with cyclic models. They do not allow cyclic evolutionary paradigms and if the universe went through a cyclic evolution such LQG models will be easily ruled out. It may seem that such regularizations then favor inflationary paradigm, but one must note that a cyclic evolution in some loop cosmological models plays an important role in setting right initial conditions for the onset of inflation \cite{hysteresis3,lqc-cyclic4}. In these ways,  unlike the standard LQC, the Thiemann regularized LQC being richer in mathematical complexity seems more likely to be phenomenologically constrained.

To conclude, let us briefly discuss some ramifications of this result in general for loop cosmologies. First it shows that there are two types of quantum gravity models allowing bouncing cosmologies. One which lead to a cyclic evolution and the other which do not. Apart from showing that cyclic evolution is not a robust prediction of all loop cosmologies, our analysis indicates that one can no longer assume that additional inputs from LQG can not dramatically change the qualitative dynamics in the post-bounce branch as understood in standard LQC. The Thiemann regularized LQC, or the mLQC-I, is an example which shows that seemingly small changes in the quantization procedure from  standard LQC may significantly alter the resulting physics. These changes are no longer hidden in the pre-bounce regime as for inflationary potentials \cite{lsw2018}. But, we should note that this does not indicate that all modifications from standard LQC which yield an asymmetric bounce would result in similar effects. As an example, a loop quantization utilizing gauge-covariant fluxes even though yielding an asymmetric bounce \cite{gauge-covariant} would still yield a viable cyclic evolution because the regularization of the constraint does not change from the one used in standard LQC.  Finally, our work  provides insights on an important question often posed in quantum gravity: How does the physics change when one imposes symmetries before the quantization (as in standard LQC) versus quantization before or without using symmetries (as in Thiemann regularized LQC)? Our analysis shows that for appropriate situations the physical effects associated with such a change in procedure can  be dramatic.

\section*{Acknowledgements}
We thank Anzhong Wang for comments on the manuscript.
This work is supported by the NSF grants PHY-1454832 and PHY-2110207, and the National Natural Science Foundation of China (NNSFC) with the Grants No. 12005186.

\end{document}